%% file: main.tex
\documentclass{article}
\input{meta}

\title{Strategic Competition and Convergence in LLM Leaderboards}

\begin{document}
\author{
\mbox{Naveen Raman\textsuperscript{1}} \thanks{%
  \protect\begin{tabular}[t]{@{}l@{}}
    Correspondence: 
    \texttt{naveenr@andrew.cmu.edu}
  \end{tabular}%
},
\mbox{Fei Fang\textsuperscript{1}},
\mbox{Yiling Chen\textsuperscript{2}},
}

\newcommand{\affiliations}{
\mbox{\textsuperscript{1}\,\href{https://cmu.edu}{Carnegie Mellon University}}\quad
\mbox{\textsuperscript{2}\,\href{https://https://www.harvard.edu/}{Harvard University}}\quad
}

\makeatletter
\renewcommand{\@maketitle}{%
  {\centering
    {\bfseries\fontsize{18}{22}\selectfont \@title\par}
    \vspace{12pt}
    {\small
      \def\@makefnmark{\hbox{\@textsuperscript{\normalfont\@thefnmark}}}%
      \@author\par}
    \vspace{6pt}
    {\footnotesize\hypersetup{urlcolor=black}\affiliations\par}%
  }
}

\maketitle

\begin{abstract}
    \input{sections/abstract}
\end{abstract}
\input{sections/introduction}
\input{sections/model}
\input{sections/equilibrium}
\input{sections/monoculture}
\input{sections/identification}
\input{sections/empirical}
\input{sections/discussion}
\input{sections/related}
\input{sections/conclusion}
\input{sections/disclosure}
\input{sections/acknowledgements}

\newpage 

\bibliographystyle{iclr2027_conference}
\bibliography{references}

\newpage 

\appendix 
\input{appendix/win_rate}
\input{appendix/no_equilibrium}
\input{appendix/experimental_details}
\input{appendix/ablations}
\input{appendix/equilibrium_proofs}
\end{document}

%% file: meta.tex
\usepackage{times}
\usepackage[letterpaper,margin=1in]{geometry}
\usepackage{iftex}
\usepackage{mathpazo}

\usepackage{natbib}

\usepackage{graphicx}
\usepackage{xcolor}

\usepackage{booktabs}

\usepackage{amsmath,amsthm,amsfonts}
\usepackage{thmtools} 
\usepackage{thm-restate}

\newtheorem{example}{Example}

\usepackage[
    colorlinks=true,
    linkcolor=blue,
    citecolor=magenta,
    urlcolor=red
]{hyperref}
\usepackage[capitalise,noabbrev]{cleveref}

\newcommand{\bx}{\mathbf{x}}
\newcommand{\bxi}{\mathbf{x}_i}
\newcommand{\bxmi}{\mathbf{x}_{-i}}
\newcommand{\btheta}{\boldsymbol{\theta}}
\newcommand{\bthetai}{\boldsymbol{\theta}_i}

\newcommand{\bw}{\mathbf{w}}

\newcommand{\BR}{\mathrm{BR}}

\newcommand{\Fi}{F_i}

\newcommand{\Hij}{H_{i,i'}}

\newcommand{\onenorm}[1]{\lVert #1\rVert_1}


%% file: sections/abstract.tex
Leaderboards are a popular way to compare large language model (LLM) capabilities and can influence how model developers invest resources.
We ask when competition on leaderboards leads developers to produce models with similar capability profiles, which we call monoculture, or similar performance levels.
We study this by formulating leaderboard competition as a game in which model developers invest resources across capabilities to maximize their win rate.
We show that similar production functions can induce similar capability profiles, while comparative advantage can sustain distinct capability profiles. 
We then study convergence in performance levels and show that diminishing returns can lead to an increasingly crowded competitive frontier.
Using data from LM Arena and OpenRouter, we find that developers improve more in categories where they lag competitors, while still retaining category-specific advantages.
At the same time, we find that the frontier of models becomes increasingly crowded, alongside increasingly diversified model usage patterns. 
Our results suggest that increasing competition does not need to produce interchangeable models, but can instead sustain competitive models with distinct strengths.

%% file: sections/introduction.tex
\section{Introduction}
\label{sec:introduction}
Leaderboards are one of the main ways in which large language models (LLMs) are measured and can influence how models are developed~\citep{benchmark_mineability}. 
While benchmark-style leaderboards remain popular, newer head-to-head leaderboards compare models based on aggregated human preferences~\citep{chatbot_arena}.
This changes strategic incentives for developers, as a model's rank depends on competitors' capabilities.

A natural question is whether this competition pushes developers toward similar performance levels, capability profiles, or both.
By providing developers with a common competitive target, shared evaluation may encourage them to invest in the same capabilities, potentially producing similar capability profiles or monoculture~\citep{monoculture_welfare}.
By contrast, differences in developer strengths across capabilities can create incentives to specialize.
At the same time, models might converge in performance levels, while still maintaining distinct capability profiles. 
Understanding this phenomenon is important because leaderboards may shape not only the models that perform best, but also the direction of model development~\citep{training_test_task,leaderboard_incentives}. 
Even if models become similarly competitive, differences in their capability profiles determine whether users have access to diverse alternatives or a homogeneous set of models. 

\begin{figure*}[t!]
    \centering 
    \includegraphics[width=\textwidth]{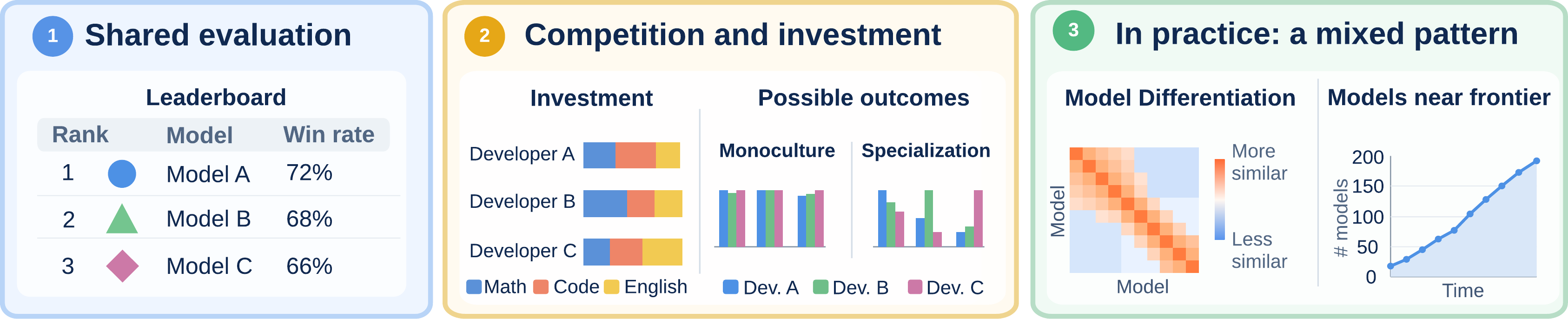}
    \caption{Head-to-head leaderboards compare models, creating incentives for developers to allocate investment across capabilities. How these incentives interact with developers' production functions can shape both the similarity of capability profiles and convergence in performance. In practice, we find persistent differences in developers' capability profiles alongside an increasingly crowded performance frontier.} 
    \label{fig:pull}
\end{figure*}

In this work, we study how leaderboard competition shapes capability profiles and performance levels by modeling AI evaluation as a multidimensional contest where developers allocate investment across capabilities (Figure~\ref{fig:pull}).
Our theoretical results characterize when similarities in production functions lead to monoculture and when diminishing returns can lead to convergence in performance.
Empirically, we find that some developers maintain differentiated capability profiles over time, alongside an increasingly crowded competitive frontier. 
Our results reveal that competition can lead to convergence in developer performance while still maintaining an ecosystem of meaningfully different models.

Our paper connects to work in contest theory, monoculture, and LLM leaderboards. 
Contest theory studies how competitors allocate effort in response to a common reward~\citep{contest_theory_book,colonel_blotto}, while algorithmic monoculture studies when shared algorithms lead to similar outcomes~\citep{monoculture_welfare,monoculture_matching}. 
LLM leaderboards combine aspects of both settings: developers compete under a common evaluation, but allocate investment across multiple capabilities. 
More recently,~\citet{leaderboard_incentives} studied strategic post-training and the resulting equilibria, while~\citet{aggregation_diversity} study how aggregation and evaluation affect producer incentives.
Unlike~\citet{leaderboard_incentives} and~\citet{aggregation_diversity}, we explicitly model how production functions shape capability development, which allows us to characterize when common evaluation produces similar capability profiles.

Our contributions include: 
\begin{enumerate}
    \item We characterize how multidimensional competition under shared evaluation produces similar or differentiated capability profiles. 
    Similar production functions can induce monoculture, while comparative advantage can sustain differentiation and, in some settings, specialization.
    \item We study how competition evolves over time and identify when diminishing returns can lead to convergence in performance. We find that convergence in performance can arise when competitors can quickly catch up in categories where they lag. 
    \item We find that the performance frontier becomes increasingly crowded over time, even as developers retain persistent differences in their relative capability profiles. Along with crowding, we find increasingly dispersed model usage across developers. \footnote{Code found at \url{https://github.com/naveenr414/evaluation_monoculture}.} 
\end{enumerate}

%% file: sections/model.tex
\section{Model and Preliminaries}
\label{sec:model}
Our model consists of $N$ model developers (e.g., OpenAI, Anthropic, DeepMind), each of which competes on $K$ dimensions (e.g., math, coding, science). 
Developer $i$ has a production function $F_{i}: \mathbb{R}_{+}^{K} \rightarrow \mathbb{R}_{+}^{K}$, and investment $\mathbf{x}_i$, which produces capabilities $\bthetai := \Fi(\bxi)$. 
Investment at level $\bxi$ results in a cost $c(\onenorm \bxi)$, where $c$ is a strictly increasing, differentiable, continuous, and unbounded cost function with $c(0) = 0$ (per-developer cost functions $c_i$ can be folded into the model without changing the equilibria; see Appendix~\ref{app:CommonCost_proof}). 
We model each developer as maximizing its expected win rate.
We use expected weighted pairwise win rate as a simplified objective for relative competitive performance, rather than assuming that developers literally optimize a particular public leaderboard.
This objective directly corresponds to maximizing weighted category-level pairwise win probabilities under Bradley–Terry/Elo-style evaluation (see Appendix~\ref{app:elo}). 
Here, win-rate is weighted according to $\bw \geq 0, \sum_{k} w_{k} = 1$.
We define the utility as 
\begin{equation}
\label{eq:utility}
u_{i}(\mathbf{x}) = \frac{1}{N-1} \sum_{i' \neq i} \sum_{k} w_{k} \sigma(\Fi(\bxi)_k - F_{i'}(\bx_{i'})_k)- c(\sum_{k} x_{i,k}).
\end{equation}
Here, $\sigma$ denotes the win-probability according to some pairwise preference model. 
We assume $\sigma$ is strictly increasing, twice differentiable, bounded in $[0,1]$, $\sigma'(x)>0$, and symmetric where $\sigma(x)+\sigma(-x)=1$ (e.g., Bradley-Terry).
Our model abstracts from the details of developers' internal R\&D processes and instead captures three features of model development: developers care about their capabilities relative to competitors, allocate costly investment across multiple capabilities, and differ in how effectively investment translates into capability improvements. 
We do not require that developers directly optimize a particular leaderboard; tournament-style evaluation provides a tractable representation of competition based on relative model performance. 
The model can also be viewed as a multidimensional contest, related to Colonel Blotto games~\citep{colonel_blotto}, in which investment is transformed into capabilities through heterogeneous production functions \(F_i\).

Let $\mathbf{x} = (\mathbf{x}_1,\ldots,\mathbf{x}_{N})$ be a Nash equilibrium, and let $\Hij(\mathbf{x}) = \cos(\bthetai,\btheta_{i'})$ be the pairwise monoculture for developers $i$ and $i'$, with $\Hij(\mathbf{x})= 1$ indicating that $\bthetai$ and $\btheta_{i'}$ are parallel. 
We can view $\btheta_i$ as consisting of a direction $\mathbf{d}^*_i \in S_{+}^{K-1}$ and a magnitude $\alpha^*_{i}$, with only the direction impacting $\Hij$. 
We focus on cosine similarity here to capture capability profiles rather than capability levels; while two developers might differ in their overall level of output, they can still be similar if their relative category outputs are similar~\citep{monoculture_outcome}. 
We return to similarity in the overall level of output in Section~\ref{sec:dynamics}. 
We define overall monoculture as $H(\mathbf{x}) = \frac{1}{{N \choose 2}} \sum_{i<i'} \Hij(\mathbf{x})$, with $H=1$ indicating that all developers have parallel $\bthetai$. 
When exactly one capability vector is zero we define $H_{i,i'}(\mathbf{x}^*)=0$, and when both are zero we define $H_{i,i'}(\mathbf{x}^*)=1$.

%% file: sections/equilibrium.tex
\section{Existence of Pure Strategy Equilibrium}
\label{sec:equilibrium}
We discuss when the competition model exhibits pure-strategy equilibrium, where we focus on pure strategies because they correspond to stable, deterministic investment profiles, which do not require developers to randomize. 
We first consider the intuition in the \(N=2,K=1\) setting, where we refer to the production function as $f$ because it is scalar. 
A sufficient condition for existence is that diminishing returns in production are sufficiently strong relative to the curvature of the evaluation function. 
That is, when $c$ is convex and $f_{i}'(x_i) > 0$, we can write this as 
\[
 \sup_z \frac{\sigma''(z)}{\sigma'(z) } < -\frac{f''_i(x_i)}{f_i'(x_i)^2} \forall x_i.
\]
The following theorem generalizes this condition to arbitrary \(N\) and \(K\).
\begin{restatable}[Equilibrium Existence]{theorem}{EquilibriumExistence} 
\label{thm:EquilibriumExistence} 
Suppose that $g_{i,k}(\mathbf{x}_i)=[F_i(\mathbf{x}_i)]_k$ is twice continuously differentiable and $c$ convex. 
Let $\rho_\sigma := \sup_{z\in\mathbb{R}}\frac{\sigma''(z)}{\sigma'(z)}$.  
There is a pure-strategy Nash equilibrium if
\[ -\nabla^2 g_{i,k}(\mathbf{x}_i) \succeq \rho_\sigma \nabla g_{i,k}(\mathbf{x}_i) \nabla g_{i,k}(\mathbf{x}_i)^\top \qquad \forall\,\mathbf{x}_i,\ \forall\,i,k. 
\] 
\end{restatable}

Equilibrium exists when developers' production functions have sufficiently strong diminishing returns relative to the curvature of the evaluation function.
Since the win-rate payoff is at most \(1\), any investment with \(c(\|\bx_i\|_1)>1\) is dominated by zero investment. 
That is, in Equation~\ref{eq:utility}, if $c(\|\bx_i\|_1)>1$ then $u_{i}(\bx_i) < 0$ because $\frac{1}{N-1} \sum_{i' \neq i} \sum_{k} w_{k} \sigma(\Fi(\bxi)_k \leq 1$. 
By unboundedness of \(c\), we can restrict each developer's strategy space to a compact convex set \(\mathcal X_i=\{\bx_i\ge0:\|\bx_i\|_1\le\bar x\}\).
Under the theorem's curvature condition, the Hessian of \(u_i\) is negative semidefinite, so \(u_i\) is concave in \(\bx_i\). Debreu–Fan–Glicksberg then gives existence (details in Appendix~\ref{app:EquilibriumExistence_proof}). 
When production functions violate the conditions, pure-strategy equilibrium need not exist (see Appendix~\ref{app:no_equilibrium}).

%% file: sections/monoculture.tex
\section{When Does Monoculture Arise?}
\label{sec:monoculture}
Given that a pure-strategy equilibrium exists, we next ask what determines whether developers produce similar capability profiles or specialize in different capabilities. 
We first give interpretable sufficient conditions for monoculture and specialization, before a general characterization.

\subsection{Similar Production Functions Can Induce Monoculture}
A simple sufficient condition for monoculture is developers having identical production functions. 

\begin{restatable}[Symmetric Developers]{lemma}{SymmetricDevelopers}
\label{lem:SymmetricDevelopers}
Suppose the conditions of Theorem~\ref{thm:EquilibriumExistence} hold. 
If $F_i=F_{i'}$, then there exists a pure-strategy Nash equilibrium $\mathbf{x}^*$ such that $\bxi^*=\bx_{i'}^*$ and $\Hij(\mathbf{x}^*) = 1$. 
Moreover, if $F_i=F$ for every developer $i$, then there exists a symmetric pure-strategy Nash equilibrium satisfying  $\mathbf{x}_1^*=\cdots=\mathbf{x}_N^*$, and at this equilibrium, $\theta_1^*=\cdots=\theta_N^*$, with $H(\mathbf{x}^*)=1$. 
\end{restatable}
Restricting attention to profiles with $\bx_i=\bx_{i'}$, the two developers have identical best-response correspondences; Kakutani's fixed-point theorem then gives the equilibrium (Appendix~\ref{app:SymmetricDevelopers_proof}).

Next, we note that strict equality is not necessary for near-monoculture. 
To establish this, we first introduce the \textit{pseudogradient} of the game, $G(\mathbf{x})$, which we define as 
\[
G(\bx)
=
\left(
-\nabla_{\bx_1}u_1(\bx),
\ldots,
-\nabla_{\bx_N}u_N(\bx)
\right).
\]
We next show that similarity in $F_i$ and $\nabla F_i$ induces similarity in outputs.
\begin{restatable}[Similarity in Production Functions]{theorem}{SimilarFHomogeneity}
\label{thm:SimilarFHomogeneity}
Suppose the conditions of Theorem~\ref{thm:EquilibriumExistence} hold, while the production functions are $L$-Lipschitz, and the pseudogradient $G$ is $\mu$-strongly monotone, which ensures equilibrium stability to perturbations:
\[
\left(G(\bx)-G(\bx')\right)^\top(\bx-\bx')
\geq
\mu\lVert\bx-\bx'\rVert_2^2
\qquad
\forall \bx,\bx'.
\]
If developers $i$ and $i'$ have similar production functions in both value and gradient:
\[
\sup_{\bx_i\in\mathcal X_i}
\left\|F_i(\bx_i)-F_{i'}(\bx_i)\right\|_2
\leq \epsilon,
\qquad
\sup_{\bx_i\in\mathcal X_i}
\left\|\nabla g_{i,k}(\bx_i)-\nabla g_{i',k}(\bx_i)\right\|_2
\leq \epsilon
\quad \forall k.
\]
then there exists an equilibrium $\bx^*$ that satisfies
\[
\lVert\bx_i^*-\bx_{i'}^*\rVert_2
\leq
O\left(\frac{\epsilon}{\mu}\right),
\qquad
\lVert F_i(\bx_i^*)-F_{i'}(\bx_{i'}^*)\rVert_2
\leq
O\left(\epsilon\left(1+\frac{L}{\mu}\right)\right).
\]

Moreover, if $\lVert\btheta_i^*\rVert_2,\lVert\btheta_{i'}^*\rVert_2\geq m>0$, then their pairwise monoculture satisfies
\[
H_{i,i'}(\mathbf{x}^*)
\geq
1-O\left(
\frac{\epsilon^2}{m^2}
\left(1+\frac{L}{\mu}\right)^2
\right).
\]
\end{restatable}

The strong-monotonicity condition holds when each developer's own investment has a sufficiently strong stabilizing effect on its incentives relative to the strategic interaction between developers. 
Therefore, the result is most applicable to settings with a stable equilibrium in which small differences in developers' production technologies do not induce large changes in equilibrium investment.
When developers have similar production functions, their equilibrium investments are similar.
Evaluation coordinates developers and makes it possible for them to optimally respond in the same way. 
The proof constructs a counterfactual game and uses strong monotonicity to translate perturbation in the pseudogradient into bounds on equilibrium outputs (Appendix~\ref{app:SimilarFHomogeneity}).

\subsection{Comparative Advantage Can Induce Specialization}
While similar production functions induce monoculture, comparative advantage, where developers' production functions favor different capabilities, can create incentives for specialization.
We first illustrate a setting where comparative advantage can induce specialization. 

\begin{example}[Comparative advantage can induce specialization]
\label{ex:comparative_advantage}
Suppose $N=K=2$, $w_1=w_2=\frac12$, and
\[
F_i([x_1,x_2])
=
[a_i\log(1+x_1),\,b_i\log(1+x_2)].
\]
The KKT conditions imply that a specialized equilibrium
$\bx_1^*=[s_1,0]$, $\bx_2^*=[0,s_2]$ must satisfy
\[
\frac{a_1b_2}{a_2b_1}
\geq
(1+s_1)(1+s_2).
\]
Such equilibria exist. 
For example, with logistic $\sigma$,
\[
a_1=b_2=1,\qquad
a_2=b_1=\frac14,\qquad
c(s)=\frac{s}{18},
\]
the profile
\[
\bx_1^*=[1,0],
\qquad
\bx_2^*=[0,1]
\]
is a Nash equilibrium. 
Moreover, the conditions of Theorem~\ref{thm:EquilibriumExistence} hold, so each developer's payoff is concave in its own investment and the KKT conditions are sufficient.
\end{example}

This example illustrates how specialization can arise when production functions are sufficiently differentiated.
We can generalize this idea by representing differences in developers' comparative advantages through differences in the marginal costs of producing capability profiles. 
Using this, we show that sufficiently different marginal capability costs imply differentiated equilibrium outputs.
When this separation cannot be explained solely by differences in overall capability levels, it also implies differentiation in capability profiles.

\begin{restatable}[Marginal Cost Differences Induce Capability Differentiation]{theorem}{SpecializedFDifferentiation}
\label{thm:SpecializedFDifferentiation}

For developer $i$, define the minimum cost of producing capability
profile $\theta$ as $C_i(\theta):= \min_{\mathbf x_i:\,F_i(\mathbf x_i)=\theta} c(\|\mathbf x_i\|_1)$.
For a collection of rival capability vectors
$Y=\{y_1,\ldots,y_{N-1}\}$, define
\[
V_Y(\theta)
:=
\frac{1}{N-1}
\sum_{y\in Y}
\sum_{k=1}^K
w_k \sigma(\theta_k-y_k).
\]

Suppose there is a convex capability region
$\Theta\subseteq\mathbb R_+^K$ containing all relevant best responses
such that, for every developer $i$ and rival environment $Y$,
\[
\phi_{i,Y}(\theta)
:=
C_i(\theta)-V_Y(\theta)
\]
has a unique interior minimizer, is $\mu$-strongly convex, and has
$L$-Lipschitz gradient on $\Theta$.

Consider developers $i\neq i'$, and suppose their marginal capability
costs satisfy
\[
\inf_{\theta\in\Theta}
\|\nabla C_i(\theta)-\nabla C_{i'}(\theta)\|_2
\ge \delta_{i,i'}>0.
\]

Let $\mathbf x^*$ be any pure-strategy Nash equilibrium, with
$\theta_i^*=F_i(\mathbf x_i^*)$ and
$\theta_{i'}^*=F_{i'}(\mathbf x_{i'}^*)$. Define
\[
\ell_\sigma:=\sup_{z\in\mathbb R}|\sigma''(z)|,
\qquad
\rho_N
:=
\frac{\ell_\sigma\|w\|_\infty}
{\mu(N-1)}.
\]
Then
\[
\|\theta_i^*-\theta_{i'}^*\|_2
\ge
\frac{\delta_{i,i'}}{L(1+\rho_N)}.
\]

Consequently, if 
$r_i=\|\theta_i^*\|_2>  0$ and $r_{i'}=\|\theta_{i'}^*\|_2 > 0$ then
\[
H_{i,i'}(\mathbf x^*)
\le
1-
\frac{
\left[
\left(
\frac{\delta_{i,i'}}{L(1+\rho_N)}
\right)^2
-(r_i-r_{i'})^2
\right]_+
}
{2r_ir_{i'}}.
\]
\end{restatable}

These conditions capture settings in which each developer has a stable capability-level best response and developers have persistent differences in their relative costs of producing capabilities. 
Such an assumption is most natural when comparative advantages arise from durable differences in production technologies, so that developers' marginal production costs remain distinct throughout the relevant range of equilibrium outputs.
When marginal capability costs are different, we can show that the resulting equilibria are also different. 
When overall capability levels are also different, we can use this to show differences in the similarity $H_{i,i'}(\mathbf{x}^*)$.
We prove this by constructing a counterfactual environment, where developer $i'$ responds to developer $i$'s environment, then bounding the difference in outputs (full proof Appendix~\ref{app:SpecializedFDifferentiation}). 

\subsection{Characterizing Monoculture and Specialization Generally}
While Theorem~\ref{thm:SimilarFHomogeneity} provides a sufficient condition, even drastically different production functions can result in monoculture.
We present such an example below:
\begin{example}
\label{ex:monoculture_not_equal}
    Consider a scenario with $N=K=2$, $\sigma(z)$ as the sigmoid function, and $c(\lVert \mathbf{x}_{i} \rVert _1) = \frac{0.15}{2} \lVert \mathbf{x}_i \rVert_1^{2}$. 
    Let $F_i(\mathbf{x}_i)=A_i\log(\mathbf{x}_i+\mathbf{1})$,
    where the logarithm is applied componentwise.
    Let 
    \[ A_1=
    \begin{pmatrix}
    0.50 & 0.05\\
    0.10 & 0.75
    \end{pmatrix},
    \qquad
    A_2=
    \begin{pmatrix}
    0.05 & 0.60\\
    0.70 & 0.15
    \end{pmatrix},
    \qquad
    \bw=(0.35,0.65).
    \]
        
    Then numerically solving for the equilibrium yields 
    \[
    \bx_1^*=(0,\ 0.545), \qquad \bx_2^*=(0.518,\ 0),
    \]
    and the corresponding outputs are 
    \[
    \btheta_1^*\approx(0.022,\ 0.326),
    \qquad
    \btheta_2^*\approx(0.021,\ 0.292).
    \]
    Here, $H_{1,2}(\mathbf{x}^*) \approx 0.99999$, and so the outputs are near-parallel while the production functions are very different. 
\end{example}

This example also demonstrates why a general characterization of monoculture is difficult, as even dissimilar production functions can result in similar capability profiles. 
Instead, we can characterize monoculture in terms
of the local production incentives that must hold at equilibrium.
We can formalize this observation using the equilibrium
first-order conditions. 
First, let $\btheta^*_{i} = \alpha^*_i d^*_{i}$.
Then for a profile of candidate directions $\mathbf D=(\mathbf d_1,\ldots,\mathbf d_N)\in(S_+^{K-1})^N$ and $\boldsymbol{\alpha}\in\mathbb R_{+}^N$, define
\[
p_{i,k}(\mathbf D,\boldsymbol{\alpha}) = \frac{w_k}{N-1} \sum_{j\neq i} \sigma'\!\left( \alpha_i d_{i,k}-\alpha_j d_{j,k} \right).
\]
For a profile of output directions, $\mathbf{D}=(\mathbf d_1,\ldots,\mathbf d_N)\in(S_+^{K-1})^N$, we say that $\mathbf{D}$ is locally supportable if there exist
$\boldsymbol{\alpha}\in\mathbb R_{+}^N$ and
$\bx_i\in\mathbb R_+^K$ such that
\[
F_i(\bx_i)=\alpha_i\mathbf{d_i}\neq 0 \forall i, \quad \text{and} \quad \nabla F_i(\bx_i)^\top
\mathbf{p}_i(\mathbf{D},\boldsymbol{\alpha})
\leq
c'(\|\bx_i\|_1)\mathbf 1 \forall i,
\]
with equality in coordinates $\ell$ when
$x_{i,\ell}>0$.
This definition essentially collects the KKT conditions relevant to the direction of each developer's output and provides a convenient way to
state exactly which capability directions can be sustained at equilibrium.

\begin{restatable}[Supportable Equilibrium Directions]
{proposition}{DirectionalCharacterization}
\label{prop:DirectionalCharacterization}
Suppose that the conditions of Theorem~\ref{thm:EquilibriumExistence} hold.
A profile of directions $\mathbf{D}=(\mathbf d_1,\ldots,\mathbf d_N)\in(S_+^{K-1})^N$
is induced by a pure-strategy Nash equilibrium with nonzero outputs if and only if $\mathbf{D}$ is locally supportable.

A pure-strategy equilibrium with monoculture exists if and only if $(\mathbf d,\ldots,\mathbf d)$ is locally supportable for some $\mathbf d\in S_+^{K-1}$, and a pure-strategy equilibrium with complete specialization exists if and only if there is a locally supportable profile $\mathbf{D}=(\mathbf d_1,\ldots,\mathbf d_N)$ with $\mathbf d_i^\top \mathbf d_{i'}=0$ for all $i \neq i'$.
\end{restatable}

Proposition~\ref{prop:DirectionalCharacterization} is a reformulation of the equilibrium first-order conditions, and demonstrates general requirements behind monoculture and specialization.
The key insight is that monoculture does not require developers to have similar production functions, but rather that their production functions need only locally support the same output direction under the competitive incentives they face
(details in Appendix~\ref{app:DirectionalCharacterization}).

Shared evaluation makes the value of a capability improvement depend not only on a developer's production technology, but also on its performance relative to competitors. 
In particular, the value of improving on capability $k$ depends on how well developer $i$ performs relative to rivals. 
Because every developer is scored against the same category weights, these incentives can align developers' investment directions even when their production functions differ, providing a second route to monoculture beyond technological similarity. 
This distinction explains why similarity in production functions is only a sufficient condition for monoculture (Lemma~\ref{lem:SymmetricDevelopers} and Theorem~\ref{thm:SimilarFHomogeneity}), whereas Proposition~\ref{prop:DirectionalCharacterization} gives a necessary-and-sufficient characterization.
That is, monoculture occurs at equilibrium exactly when developers' production technologies support a common capability direction under these competitive marginal incentives.
We next tackle whether performance levels converge over time even when developers retain different relative strengths across capabilities.

%% file: sections/identification.tex
\section{Competitive Dynamics}
\label{sec:dynamics}
Section~\ref{sec:monoculture} characterized when developers' equilibrium capability profiles exhibit monoculture or specialization. 
We now turn from the direction of capabilities at equilibrium to the evolution of performance levels over time. 
In particular, we ask when competing developers become closer in performance as investment accumulates, even if differences in their capability profiles persist.

Let $\mathbf{X}_{i,t} = \sum_{s=1}^t \mathbf{x}_{i,s}$
denote developer $i$'s cumulative investment through time $t$.
Additionally, let $\boldsymbol{\theta}_i^t = F_i(\mathbf X_i^t)$ denote developer $i$'s capability vector at time $t$.
At each time step $t$, each developer chooses $\mathbf{x}_{i,t}$ to maximize its current-period utility:
\[
\frac{1}{N-1}
\sum_{i'\neq i}\sum_k
w_k
\sigma\!\left(
F_i(\mathbf{X}_{i,t-1}+\mathbf{x}_{i,t})_k
-
F_{i'}(\mathbf{X}_{i',t-1}+\mathbf{x}_{i',t})_k
\right)
-
c(\|\mathbf{x}_{i,t}\|_1).
\]
This abstraction isolates how repeated investment changes relative performance without introducing intertemporal strategic considerations. 
We ask whether developers’ relative performance converges over time.
We do not expect competition to guarantee convergence: developers may remain separated because their cumulative investments diverge or because differences in their production functions persist at scale.
Rather than characterize the full dynamic equilibrium, we isolate conditions on investment trajectories and production functions under which relative performance must converge.

\begin{restatable}[Catch-Up and Long-Run Crowding]
{proposition}{AsymptoticBunching}
\label{prop:AsymptoticBunching}

Let $\boldsymbol{\theta}_i^t = F_i(\mathbf{X}_i^t)$ denote developer $i$'s capability vector at time $t$, and define its category-specific gap to the frontier as $g_{i,k}^t=\max_j \theta_{j,k}^t-\theta_{i,k}^t.$
Let $\mathbf g^t\in\mathbb R_+^{NK}$ stack these gaps across developers and
categories at timestep $t$, and let
\[
\Delta\boldsymbol{\theta}^t
=
\left(
\boldsymbol{\theta}_1^{t+1}-\boldsymbol{\theta}_1^t,
\ldots,
\boldsymbol{\theta}_N^{t+1}-\boldsymbol{\theta}_N^t
\right)
\in\mathbb R^{NK}.
\]

Suppose each $F_i$ is continuously differentiable with
\[
\max_{\ell\in[K]} \left\| \frac{\partial F_i(\mathbf x)}{\partial x_\ell} \right\|_2 \to 0 \qquad \text{as }\|\mathbf x\|_1\to\infty, \quad \forall i.
\]
Suppose there exists $\beta>0$ such that, for sufficiently large $t$, $\left\langle
\mathbf g^t,\Delta\boldsymbol{\theta}^t
\right\rangle
\ge
\beta\|\mathbf g^t\|_2^2$.
Then we have that $\|\mathbf g^t\|_2\to 0$, and $\max_{i'}\theta_{{i'},k}^t-\theta_{i,k}^t
\to 0$ for all developers $i$ and categories $k$. 
\end{restatable}

The competitive dynamics generate a sequence of relative performance gaps and per-period investments that decrease over time, while the catch-up condition requires improvements to be directed toward those gaps (proof in Appendix~\ref{app:AsymptoticBunching}).

We next give one set of sufficient conditions under which diminishing returns and competitive incentives generate catch-up. 
The key additional requirement is that a developer that trails in a capability retains marginal capacity to improve that capability at a rate related to the size of its gap.
Intuitively, lagging capabilities retain greater marginal capacity for improvement, while competition creates incentives for developers to perform well in lagging capabilities.

\begin{restatable}[Sufficient conditions for capability catch-up]{proposition}{PrimitiveCatchup}
\label{prop:PrimitiveCatchup}
Suppose each $F_i$ is continuously differentiable and coordinatewise
nondecreasing, each coordinate function $[F_i(\cdot)]_k$ is concave, and
$F_i$ is $L$-Lipschitz with respect to $\ell_1$.
Additionally, suppose $w_k>0$ with $w_{\mathrm{min}} = \min\limits_{k \in [K]} w_k$, and $c$ is twice continuously differentiable with $c'(0)=0$ and $0\le c''(s)\le M$. 
Assume that each developer is playing a pure-strategy Nash equilibrium $\mathbf{x}^*_{i,t}$ at each timestep.  

For each developer $i$, category $k$, and investment state $\mathbf x$, define
\[
    m_{i,k}(\mathbf x)
    :=
    \min_{\ell\in[K]}
    \frac{\partial [F_i(\mathbf x)]_k}{\partial x_\ell}.
\]
Suppose there exists $\mu>0$ such that, for every pair of developers
$i,i'$, category $k$, and relevant investment states $\mathbf x,\mathbf y$,
whenever $[F_i(\mathbf x)]_k \le [F_{i'}(\mathbf y)]_k$, we have
\[ 
    m_{i,k}(\mathbf x)^2
    \ge
    \mu\left(
        [F_{i'}(\mathbf y)]_k-[F_i(\mathbf x)]_k
    \right).
\] 
Define
\[
    \underline p
    :=
    w_{\mathrm{min}}
    \min_{|z|\le \frac{L^2}{\mu}}\sigma'(z)>0.
\]
Then, for every developer $i$, category $k$, and period $t$,
\[
    \Delta\theta_{i,k}^t
    \ge
    \beta g_{i,k}^t,
    \qquad
    \beta
    :=
    \frac{\underline p\,\mu}
         {M+\underline p\,\mu}
    >0.
\] 
\end{restatable}

Proposition~\ref{prop:PrimitiveCatchup} illustrates how diminishing returns and competitive incentives can jointly generate catch-up. 
The parameter \(\mu\) captures how much marginal production capacity a developer retains relative to its capability gap, while \(\underline{p}\) lower-bounds the competitive value of improving a capability. 
As a result, the catch-up guarantee is stronger when lagging capabilities retain greater marginal production capacity, competitive evaluation places greater marginal value on improvement, and marginal investment costs grow more slowly. 
The diminishing returns assumption is plausible when capability gaps are bounded and lagging developers have not exhausted the channels available for improving that capability; it becomes restrictive when some investment inputs have little or no effect on a capability or when large gaps coexist with very small marginal returns. 
Diminishing returns allows lagging capabilities greater remaining capacity for improvement, while competition gives developers an incentive to exploit this capacity. 

The proof combines the lower bound on marginal production capacity with the developers' first-order conditions to obtain a lower bound on the capability improvement of a trailing developer (Appendix~\ref{app:PrimitiveCatchup}).
Because the cost function aggregates investment across dimensions, the equilibrium conditions determine the total amount of investment but not, absent solving the equilibrium, how that investment is distributed across inputs. 
A uniform catch-up guarantee therefore requires a lower bound that holds for every possible active investment direction, which leads naturally to the minimum marginal effect across dimensions.
Relaxing this assumption therefore requires a different cost function that places more structure on how investment is allocated across dimensions, or additional assumptions about which investment dimensions are active at equilibrium. 
As a result, this condition should be interpreted as a sufficient primitive condition for catch-up, rather than as a necessary condition for catch-up.

Sections~\ref{sec:monoculture} and~\ref{sec:dynamics} characterize two dimensions of convergence. 
Developers may become similar in their capability profiles (monoculture), or in their overall performance levels (crowding). 
Our empirical analysis asks how these two forms of convergence evolve over a finite period, and whether frontier crowding can coexist with persistent differences in relative capability profiles. 

%% file: sections/empirical.tex
\section{Leaderboard Competition in Practice}
We now investigate whether capability profiles are differentiated in practice and if performance converges over time. 
We first study capability profile differentiation by looking at whether developers exhibit distinct strengths (RQ1) and whether developers improve more in categories  where they lag competitors (RQ2). 
We then study convergence in performance by examining whether the competitive frontier becomes more crowded over time (RQ3). 
Together, these analyses ask whether the patterns generated by our theoretical model appear in observed model competition rather than directly testing the assumptions of our theoretical model. 

\subsection{Dataset Details}
We rely on two sources of data: leaderboards from LM Arena and usage data from OpenRouter.
\paragraph{Leaderboards}
We use LM Arena to construct leaderboards and better understand how capability profiles evolve over time. 
We construct a panel covering 422 models from 59 organizations and 28 categories, using the ``lmarena-ai/leaderboard-dataset'' from March 2024 through September 2026. 
Each snapshot contains 28 category-level Elo ratings, along with rating variances and vote counts. 
Of these 28 categories, we focus on 8 (coding, math, hard prompts, instruction following, creative writing, longer query, multi-turn, and English) because these categories are consistently available since 2024. 

\paragraph{OpenRouter}
To measure realized model usage, we use daily OpenRouter traffic data from January 2025 through September 2026. The data cover 408 models over 620 days.
We aggregate token shares to the model and developer levels to study how changes in the evaluated model frontier correspond to changes in realized model usage.
We detail how we match data from OpenRouter up with data from LM Arena in Appendix~\ref{app:matching_details}. 

\subsection{RQ1: Do developers retain distinct capability profiles?}
\begin{figure*}
    \includegraphics[width=0.657\textwidth]{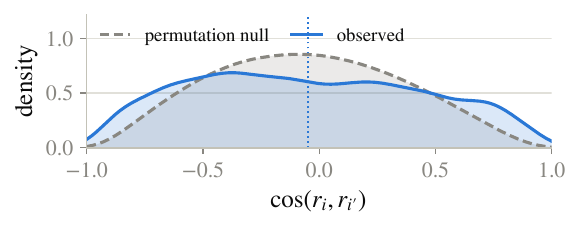}
    \includegraphics[width=0.343\textwidth]{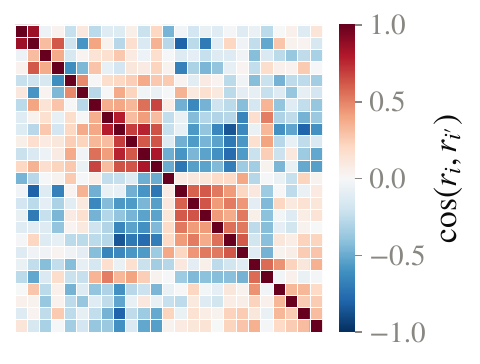}
    \caption{Developers exhibit distinct capability profiles overall, with clusters of similar relative strengths. At each snapshot, we represent each developer by its highest-rated model and compare residualized capability profiles. Left: distribution of pairwise cosine similarities across developers, compared with a permutation null that randomly reassigns capabilities across developers. Right: pairwise similarities reveal clusters of developers with similar relative strengths.}
    \label{fig:residual_cosine_kde}
\end{figure*}

We first examine whether developers' capability profiles are differentiated, motivated by Section~\ref{sec:monoculture}. 
Our goal is to isolate which categories each developer performs relatively well on, accounting for both category difficulty  (e.g., math being easier than English) and dominant developers (e.g., Anthropic being broadly better). 
However, directly applying the metrics from Section~\ref{sec:monoculture} to Elo scores is inappropriate because Elo scores are identified only up to an additive shift. 
Instead, at each timestep, we select the highest rated model for each developer and compare capability profiles. 
To account for differences in overall developer strength and category difficulty, we fit $\theta_{i,k} = \delta + \eta_i + \gamma_k + r_{i,k}$ for each time step separately. 
Here, differences in $r_{j,k}$ capture differences in capability profiles, while $\eta_i$ captures differences in overall performance. 
This metric captures similarities in the capability profiles for different developers, rather than directly measuring the theoretical cosine similarity from Section~\ref{sec:monoculture}.  

In Figure~\ref{fig:residual_cosine_kde} (left), we plot the distribution of $\cos(r_{i,\cdot},r_{i',\cdot})$ between pairs of developers.
We find little evidence of a common capability profile across all developers, as the observed similarities are close to a null model (see Appendix~\ref{app:statistical_rq1} for details and statistical tests). 
However, in Figure~\ref{fig:residual_cosine_kde} (right), we do find that clusters of developers tend to be similar. 
For example, models from Google and Meta are stronger in creative writing and multi-turn interactions and weaker in coding and math. 
These clusters cut across developer types; for example, models from OpenAI and Anthropic are in a different cluster than those from Google and Meta. 

\begin{figure*}[t]
    \includegraphics[width=0.5\textwidth]{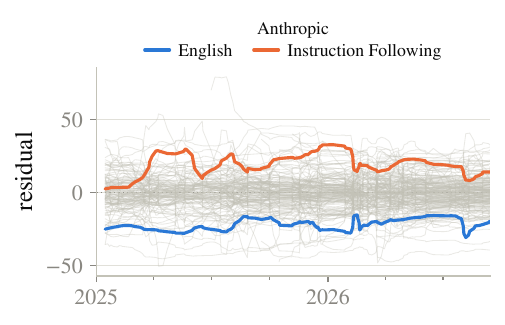}
    \includegraphics[width=0.5\textwidth]{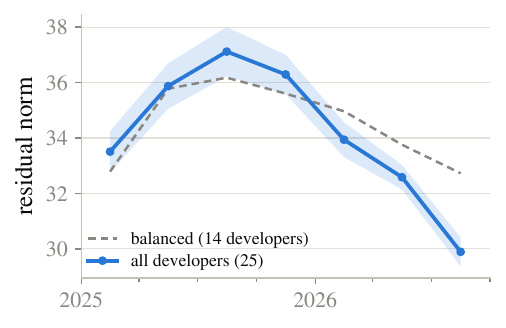}
    \caption{Some developer-specific capability advantages persist over time, even as the overall magnitude of specialization changes. Left: most developer-category residuals are noisy, but a few persist across time. For example, Anthropic's score in English and its score in instruction following hold in sign across time steps. Right: magnitude of residuals over time.}
    \label{fig:residual_distribution}
\end{figure*}

We next ask whether these relative strengths persist over time. 
Figure~\ref{fig:residual_distribution} (left) shows that many developer--category residuals fluctuate around zero, but some exhibit stable advantages or disadvantages. 
In particular, we find that 29 of the 104 developer--category pairs have the same sign in every quarter. 
For example, Anthropic persistently underperforms in English, while Cohere overperforms in creative writing. 
Figure~\ref{fig:residual_distribution} (right) plots the  L2 norm of each developer's residual vector, \(\lVert r_{i,\cdot,t} \rVert_2\), which captures the magnitude of category-specific specialization. 
We see that this magnitude first increases then decreases over time, indicating that developers' capability profiles become less specialized after a certain time. 
This pattern holds both for the 14 developers that appear in every quarter and the full set of developers, including those entering in 2025 and 2026. 
Together, these results suggest that developers retain some persistent differences in their relative strengths, even as the overall magnitude of specialization decreases. 
While the magnitude of specialization decreases in later periods, we find little evidence of a uniform capability monoculture.

\subsection{RQ2: Do developers improve more in capabilities where they lag competitors?}
We investigate how differences in capability profiles relate to subsequent improvement by looking at how developers improve or catch up to competitors. 
This examines an empirical analogue of the catch-up condition in Proposition~\ref{prop:AsymptoticBunching}.
Specifically, we relate a developer's improvement during new model releases, \(\Delta\theta_{i,k,t}\), to its pre-release gap relative to the best rival \(g_{i,k,t} = \theta_{\mathrm{leader},k,t} - \theta_{i,k,t}\).
We fit $\Delta\theta_{i,k,t} = \eta_{i,k} + \gamma_{k,t} + \beta g_{i,k,t} + \varepsilon_{i,k,t}$, where $t$ is the calendar quarter of the release, and $\eta$ and $\gamma$ represent fixed effects (details in Appendix~\ref{app:statistical_rq2}). 
We estimate $\hat{\beta}$, which captures whether larger competitive gaps predict subsequent improvement, for the 12 developers with sufficiently many model releases (more details in Appendix~\ref{app:statistical_rq2}).

\begin{figure*}[t]
    \includegraphics[width=0.5\textwidth]{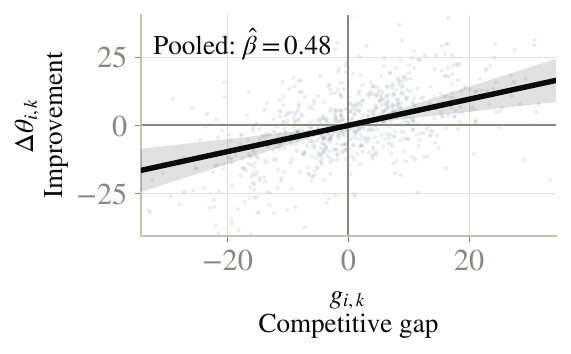}
    \includegraphics[width=0.5\textwidth]{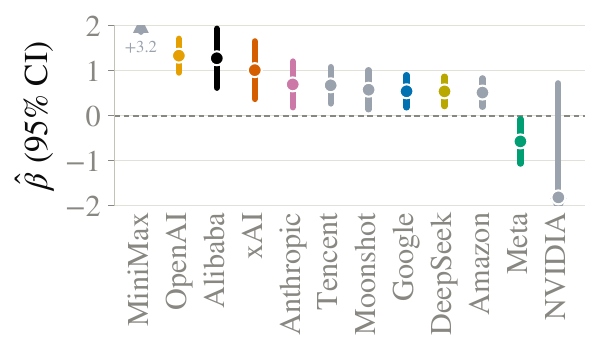}
    \caption{Developers catch up more in categories where they lag further behind the leader. Left: relationship between competitive gaps and subsequent improvements, where $\hat{\beta}$ denotes the slope. Right: developer-specific estimates of the catch-up coefficient, with 95\% confidence intervals.}
    \label{fig:evalchasing_scatter}
\end{figure*}

Figure~\ref{fig:evalchasing_scatter} shows that developers catch up more in categories where they lagged the frontier, with an estimated coefficient of $\hat{\beta}=0.48$. 
Intuitively, a 10-point larger competitive gap is associated with 4.8 points more subsequent improvement, conditional on the fixed effects.
This relationship holds across most developers, as $\hat{\beta}>0$ for 10 out of 12 developers, with the 95\% confidence interval excluding zero. 
Here, the standard error is similar when clustering by either release event ($0.117$) or developer ($0.113$).

\begin{figure*}[t]
    \centering 
    \includegraphics[width=\textwidth]{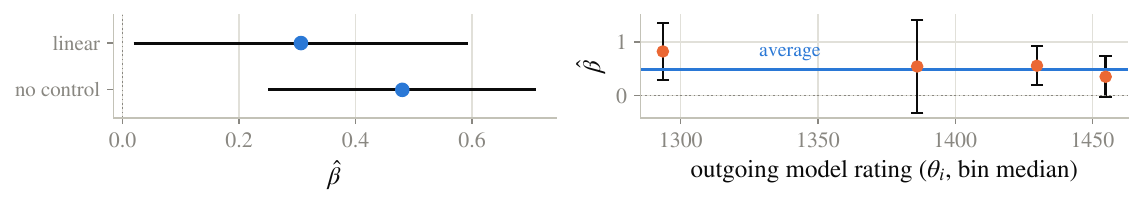}
    \caption{The catch-up relationship remains positive after controlling for overall model strength. Left: estimates of \(\hat{\beta}\) after adding a linear control for overall model strength, \(\theta_{i,\cdot}\). Right: estimates of \(\hat{\beta}\) across quartiles of overall model strength.}
    \label{fig:score_control}
\end{figure*}

To better understand the factors underlying this pattern, we investigate whether this improvement reflects differences in overall model strength, developers' category-specific weaknesses, or the strength of competitors. 
Figure~\ref{fig:score_control} accounts for overall baseline model performance, \(\theta_{i,\cdot}\), in two ways: by directly controlling for \(\theta_{i,\cdot}\) in the regression (left), and by estimating \(\hat{\beta}\) separately across quartiles of \(\theta_{i,\cdot}\) (right).
In both cases, \(\hat{\beta}\) remains positive, though the catch-up relationship is largest among models with lower baseline performance.
We additionally decompose the frontier gap into a developer's own category-specific performance and the performance of its strongest rival in Appendix~\ref{app:strong_competitors}.
Our results suggest that developers particularly catch up in categories where they themselves are weak, rather than simply in categories with unusually strong rivals.
Together, these results suggest that the observed improvement pattern is primarily associated with developers' category-specific weaknesses, rather than their overall performance or unusually strong competitors.

We conduct additional robustness checks in Appendix~\ref{app:confounds} by accounting for release year, regression to the mean, and comparator choice. 
The positive relationship persists across alternative specifications, while our regression-to-the-mean analysis suggests that measurement noise alone does not explain the observed association.
Together, RQ1 and RQ2 reveal that developers retain distinct relative strengths across capabilities, even as they tend to improve more in areas where they lag their competitors. 
We next examine how performance gaps between models have changed over time. 

\subsection{RQ3: Are models converging in performance?}
\begin{figure*}[t]
    \includegraphics[width=\textwidth]{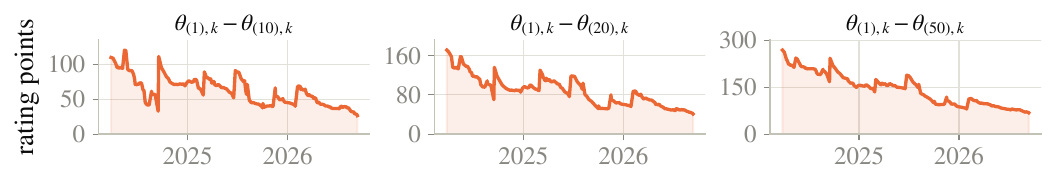}
    \caption{Performance gaps narrow across the competitive frontier. We measure gaps between the leader and the 10th-, 20th-, and 50th-ranked models, averaged across categories.}
    \label{fig:near_top}
\end{figure*}

\begin{figure*}[t]
  \includegraphics[width=0.33\textwidth]{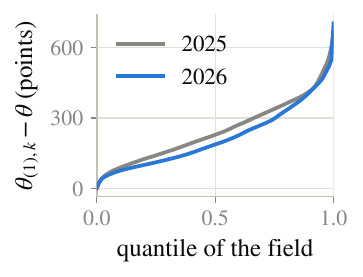}
  \includegraphics[width=0.33\textwidth]{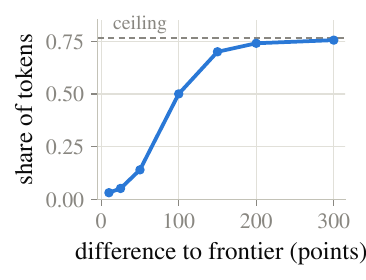}
  \includegraphics[width=0.33\textwidth]{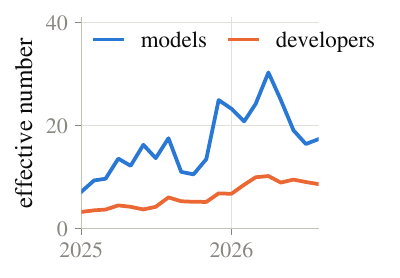}
    \caption{Models become closer in performance over time, alongside increasingly dispersed usage. Left: models in 2026 lie closer to the performance frontier than in earlier years. Middle: OpenRouter token usage is concentrated among models near the Arena leader. Right: the effective number of models and developers used on OpenRouter has increased over time.}
    \label{fig:top_distro}
\end{figure*}

Having found that developers tend to improve more in capabilities where they lag competitors (RQ2), we next examine whether performance gaps between competing models have narrowed over time.
We do so by measuring the performance gap between the leader and lower-ranked models, which is the empirical analog of $\mathbf{g}^{t}$ from Proposition~\ref{prop:AsymptoticBunching}, averaged across categories. 
Figure~\ref{fig:near_top} shows increasing crowding; between July 2024 and September 2026, the absolute gaps between the leader and the 10th-, 20th-, and 50th-ranked models decrease by 53\%, 61\%, and 66\%, respectively.  
This pattern is not driven solely by developers releasing multiple models: the number of distinct developers within 100 and 200 points of the leader increases by 2.4$\times$ and 1.4$\times$, respectively. 
We further investigate this in Figure~\ref{fig:top_distro} (left), where we show that models in 2026 lie closer to the frontier than models from 2025. 
These results demonstrate performance convergence for an increasingly large set of models over time. 

We next investigate whether this crowded frontier is accompanied by changes in model usage. 
We do so by matching data between Arena and OpenRouter, covering 71.3\% of overall OpenRouter token volume, and 76.7\% average monthly token volume (see Appendix~\ref{app:matching_details}).
Figure~\ref{fig:top_distro} shows that OpenRouter usage is concentrated among near-frontier models; models within 150 points of the Arena leader account for 70\% of observed tokens. 
At the same time, OpenRouter usage is becoming increasingly dispersed across models and developers. 
Between January 2025 and August 2026, the effective number of models and developers--defined as the inverse Herfindahl index of token shares--used on OpenRouter increases by 2.4\(\times\) and 2.7\(\times\), respectively.
Increasing competition near the performance frontier coincides with a broader distribution of realized model usage.
Together, our empirical results reveal convergence in performance without uniform monoculture in capability profiles. 
Developers retain persistent differences in their relative strengths (RQ1), even as improvements are concentrated in capabilities where they lag competitors (RQ2) and the performance frontier becomes increasingly crowded (RQ3). 

%% file: sections/discussion.tex
\section{Discussion}
\paragraph{Competition without monoculture}
Our results demonstrate that developers can converge in performance while still maintaining distinct relative strengths.
At the same time, we observe an increasingly crowded performance frontier alongside diversified model usage. 
These findings suggest a potentially encouraging future for AI competition: rather than converging to a single dominant model or a set of interchangeable alternatives, the ecosystem may support many similarly competitive models with distinct strengths. 
If these differences translate into meaningful advantages for different users and tasks, greater competition could expand effective choice rather than eliminate it.

\paragraph{What role does evaluation play?}
Public evaluations provide a common signal of which capabilities are rewarded and how developers compare, giving competitors a shared target toward which to optimize.
Without a common signal, developers might face more dispersed incentives, potentially encouraging differentiation or specialization~\citep{competition_diversity}.
While developers' production functions determine which capabilities are easier for them to improve, shared evaluation shapes which improvements are strategically valuable, as their returns depend on their performance relative to competitors.
Even when developers face identical evaluation incentives, differences in their production functions can sustain differentiated capability profiles, rather than inevitably producing monoculture.

\paragraph{What role do evaluators play?}
A natural question is the role evaluators play in shaping competitive dynamics.
In our model, evaluators choose $\bw$, which impacts developer incentives.
Evaluators might be able to influence where competitive pressure is concentrated within our model.
For example, increasing the weight on a capability makes improving on it more valuable to developers, all else equal.
Evaluation design can therefore shape not only how model performance is measured, but also which capabilities developers prioritize and, ultimately, the competitive landscape itself.

\paragraph{Limitations}
Our theoretical model abstracts features of real-world competition.
We assume developers optimize expected win rate rather than rank, which simplifies the strategic environment and affects equilibrium existence.
In Appendix~\ref{app:elo} we show that this objective directly captures weighted category-level pairwise win probabilities under Bradley--Terry/Elo-style evaluation.
Our theoretical results characterize mechanisms through which evaluation can shape competitive investment, while our empirical results document related patterns in practice; they do not establish that evaluation causes the observed patterns of convergence. 
Causal attribution is particularly difficult in this setting because competition and model development evolve jointly.
Moreover, differences in evaluated capability profiles do not necessarily translate into meaningful differences in downstream performance or user outcomes.
Finally, our usage results rely on OpenRouter, which captures only a subset of real-world model usage and includes traffic routed through automatic model-selection mechanisms.

%% file: sections/related.tex
\section{Related Work}
\paragraph{Leaderboards}
Our work studies how preference-based leaderboards coordinate firms’ incentives and, in turn, shape diversity.
Leaderboards crowdsource language-model preferences and provide granular information about preference~\citep{chatbot_arena,copilot_arena}.
Designing such leaderboards raises at least two challenges: aggregating heterogeneous preferences into a representative ranking~\citep{informative_leaderboards} and making that ranking robust to strategic manipulation. 
The first challenge is closely related to estimating top-$k$ rankings from limited pairwise-comparison data~\citep{spectral_mle,condorcet_pairwise}, and can be viewed as a social choice problem~\citep{pluralistic_leaderboards,evaluation_social_choice}.
For the second challenge, prior work has shown that leaderboards can be manipulated through vote rigging~\citep{vote_rigging}, strategic candidacy~\citep{leaderboard_illusion,strategic_candidacy}, and targeted manipulation~\citep{dropping_preferences}.
Closest to our theoretical setting is \citet{leaderboard_incentives}, which studies how benchmark design induces strategic post-training and asks whether the resulting equilibrium ranks models according to their latent quality. 
We instead ask how heterogeneity in developers' production technologies interacts with a fixed common evaluation. 
We model costly investment through heterogeneous production functions, which allows us to characterize when production-side comparative advantage sustains specialization, and to study the dynamics of catch-up and frontier crowding.

\paragraph{Contest Theory}
Contest theory studies how strategic agents expend costly effort to improve their probability of success~\citep{borel_contest_theory,advertising_contest,contest_success_function}. 
Our model is closely related to multi-battle contests~\citep{tullock_contest,multi_battle_contests}, in which agents allocate costly resources across multiple competitive dimensions. 
A key difference is that we separate investment from competitive performance: developer-specific production functions \(F_i\) map multidimensional investment into capability profiles, which are then compared through the contest. 
This allows us to study how heterogeneity in production technologies interacts with a common evaluation signal to determine whether equilibrium outputs become similar or specialized, rather than focusing primarily on effort allocation or contest/prize design.

\paragraph{Strategic Machine Learning}
In many real-world settings, agents respond strategically to machine learning algorithms, as deploying prediction algorithms might either impact agent responses for future predictions themselves~\citep{strategic_classification,strategic_investment,performative_prediction,strategic_ranking}.
These strategic responses can also impose social costs, for example by inducing agents to expend additional effort simply to maintain favorable classifications~\citep{social_cost_strategic_classification}.
Our setting reverses the usual direction of strategic response: rather than individuals adapting to a deployed model, model developers adapt their investments to an evaluation mechanism.

\paragraph{Competition and Monoculture}
Finally, our work connects to a growing literature on algorithmic monoculture and the effects of competition on diversity.
Algorithmic monoculture describes settings in which multiple decision makers rely on similar algorithms, rankings, or predictions, causing their decisions to become correlated~\citep{correlated_errors,subjectivity_monoculture,monoculture_outcome}.
Prior work has shown that competition can act as an anti-coordination force, encouraging developers to differentiate~\citep{competition_diversity}.
We identify a countervailing force; when competitors are evaluated against a common benchmark, evaluation can coordinate investment toward the same capabilities.
Related work similarly shows that competition among increasingly capable models need not increase diversity and can, in some settings, reduce welfare~\citep{bayes_risk_welfare}.
Closest to our diversity question is \citet{aggregation_diversity}, which studies how aggregation and evaluation objectives affect producer incentives and consumer utility, showing that standard win rate can encourage homogenization while alternative weighted win-rate objectives can encourage specialization. 
We instead model the underlying investment process explicitly: developers allocate costly resources across capabilities through heterogeneous production functions. 
This allows us to characterize how common evaluation interacts with production-side comparative advantage to generate monoculture or specialization, and to study the dynamics of catch-up and frontier crowding.

%% file: sections/conclusion.tex
\section{Conclusion}
Leaderboards not only measure model capabilities, but can also shape how developers allocate resources. 
Our model shows that this competition can lead to monoculture or specialization depending on the underlying production functions.
Similar production functions can induce monoculture, while comparative advantage can induce differentiated capability profiles and, in some settings, specialization.
We additionally characterize conditions under which diminishing returns and catch-up lead to convergence in performance over time.
In practice, these forces coexist: developers improve more where they lag while retaining persistent category-specific differences, alongside an increasingly crowded model frontier.
Our findings highlight that convergence in competitive performance need not imply monoculture in capabilities, leaving room for a future with many similarly competitive but differentiated models.

%% file: sections/disclosure.tex


%% file: sections/acknowledgements.tex
\section*{Acknowledgements}
We would like to thank Gaurav Ghosal, Kanad Pardeshi, Kevin Li, Aishik Nagar, and Wayne Chi for their comments on this paper. 
We would additionally like to thank Ariel Procaccia, David Parkes, Manish Raghavan, Serena Wang, Kate Donahue, and Itai Shapira for productive discussions. 
We also want to thank the machine learning foundations group at Harvard for their comments on an earlier version of this work. 
Co-author Raman is supported by an NSF GRFP Fellowship and a K\&L Gates Fellowship.

%% file: appendix/win_rate.tex
\section{Win Rate and Elo}
\label{app:elo}

We discuss the relationship between the win-rate objective in our model and Bradley--Terry/Elo-style rating systems.
Recall that developer $i$'s objective depends on its pairwise probability of beating other developers across categories.
In category $k$, the probability that developer $i$ beats developer $i'$ is $\sigma\left(
F_{i}(\mathbf{x}_i)_k-F_{i'}(\mathbf{x}_{i'})_k
\right)$.
When $\sigma$ is logistic, this corresponds exactly to a Bradley--Terry model within each category, where $F_{i}(\mathbf{x}_i)_k$ is the latent performance of developer $i$ in category $k$.
Elo uses the same logistic relationship between differences in latent scores and pairwise win probabilities, up to a rescaling of the scores.

Our model aggregates these pairwise win probabilities across categories.
For category weights $\mathbf{w}$, the overall win probability of developer $i$ against developer $i'$ is
\[
\sum_{k=1}^K w_k
\sigma\left(
F_{i}(\mathbf{x}_i)_k-F_{i'}(\mathbf{x}_{i'})_k
\right).
\]
Thus, the objective in our model corresponds to developers maximizing a linear function of their category-level win rates.
This directly captures settings in which models are evaluated according to overall win rate, or according to a weighted aggregation of category-level performance.

This is distinct from fitting a single latent Bradley--Terry rating to each model using comparisons pooled across categories.
In general,
\[
\sum_{k=1}^K w_k \sigma(z_k)
\neq
\sigma\left(\sum_{k=1}^K w_k z_k\right),
\]
since a weighted average of logistic functions is not itself generally logistic unless $z_k$ is constant.
Thus, when production and performance are inherently category-specific, there need not exist a single Bradley--Terry score whose implied pairwise probabilities coincide with the weighted win probabilities in our model.
A leaderboard based on a single Bradley--Terry/Elo score therefore imposes an additional one-dimensional representation of model performance.
Our model instead directly represents category-level win probabilities and developers' incentives to improve their weighted aggregate win rate.

%% file: appendix/no_equilibrium.tex
\section{Situations Where an Equilibrium Doesn't Exist}
\label{app:no_equilibrium}
We note that an equilibrium is not guaranteed to exist if $\sigma$ is close to the step function. 
For example, suppose that 
\[
f_i(x_i)=x_i, \qquad c(x_i)=\alpha x_i, \qquad \sigma_\tau(z)=\frac{1}{1+e^{-\tau z}}, \qquad \tau > 4 \alpha.
\]
Here, $u_i(x_i;x_j)=\sigma_\tau(x_i-x_j)-\alpha x_i$, where $\tau$ controls how sharply the evaluator distinguishes between developers; as $\tau\rightarrow\infty$, $\sigma_\tau$ approaches a step function.

We first show that no symmetric candidate equilibrium, $x_1=x_2=x$ exists when $\tau  > 4\alpha$. 
At such a profile,
\[ 
\frac{\partial u_i}{\partial x_i}=\sigma_\tau'(0)-\alpha=\frac{\tau}{4}-\alpha > 0.
\] 
Thus, holding the opposing developer's investment fixed, each developer can increase its utility by slightly increasing its own investment. 
We note that investments above $\bar{x}$ cannot be optimal, and so no symmetric profile can be a Nash equilibrium.
This is because, beyond $\bar{x}$, each developer achieves negative utility, and so would prefer to deviate.

Next, we demonstrate that an asymmetric equilibrium cannot exist either. 
Suppose without loss of generality that $x_1>x_2$. 
When a developer is behind, we have \(\sigma_\tau''(x_2-x_1)>0\), so its utility is strictly convex in its own investment. 
As a result, the losing developer cannot have an interior optimum strictly below its competitor. 
Therefore, any asymmetric equilibrium must, up to relabeling, take the form \((x,0)\) for some \(x>0\).
Since $x$ is optimal for the winning developer against zero, it satisfies
\[
u_1'(x;0)=\sigma_\tau'(x)-\alpha=0.
\]
Moreover, since $\tau>4\alpha$, we have $u_1'(0;0)=\tau/4-\alpha>0$. 
Since $\sigma_\tau'(y)$ is strictly decreasing for $y>0$, it follows that $u_1'(y;0)>0$ for every $y\in[0,x)$. Hence
\[
u_1(x;0)=\sigma_\tau(x)-\alpha x>u_1(0;0)=\frac12.
\]
We now show that such a solution cannot be an equilibrium. 
If developer $2$ deviates from $0$ to $2x$, its gain is
\[
u_2(2x;x)-u_2(0;x)
=\sigma_\tau(x)-2\alpha x-\sigma_\tau(-x)
=2\left(\sigma_\tau(x)-\alpha x-\frac12\right)>0,
\]
where the final inequality follows from \(u_1(x;0)>1/2\). Thus, developer 2 strictly prefers to leapfrog developer 1, so \((x,0)\) cannot be an equilibrium. Therefore, no pure-strategy Nash equilibrium exists.

%% file: appendix/experimental_details.tex
\section{Experimental Details}
\label{app:experimental_details}

\subsection{Name Matching Details}
\label{app:matching_details}

For our usage analysis, we match model names between OpenRouter and LM Arena, which use different naming conventions for provider prefixes, routing variants, release dates, and model versions. 
We first canonicalize names by removing provider and routing information and matching models based on their alphabetic and numeric components.
We then allow matches that differ only in date stamps, and manually verify remaining models with material token volume. 
We deliberately avoid general fuzzy matching, since incorrect matches would attribute token volume to the wrong capability profile. 
Overall, we match 213 of 408 OpenRouter models, covering 71.3\% of pooled token volume and an average of 76.7\% of monthly token volume.
Models without a defensible LMArena counterpart, including pre-release codenames and models not separately evaluated by LMArena, are left unmatched. 
We treat unmatched token volume as falling outside any threshold of the top-rated model, making our estimates of usage near the frontier conservative.

\subsection{Noise Correction}
\label{app:noise_correction}
The leaderboard reports an Elo rating \(\hat\theta_{ikt}=\theta_{ikt}+\varepsilon_{ikt}\), together with an estimated sampling variance \(v_{ikt}=\operatorname{Var}(\varepsilon_{ikt})\).
Because the outgoing model's rating enters both sides of our regression, sampling noise mechanically induces a positive relationship between the measured gap and subsequent improvement. 
In particular,
\[
\Delta\hat\theta_{ik} = \hat\theta^{\mathrm{new}}_{ik} - \hat\theta^{\mathrm{old}}_{ik}, \qquad \hat g_{ik} = \hat\theta^{\mathrm{riv}}_{ik} - \hat\theta^{\mathrm{old}}_{ik},
\]
so \(\varepsilon^{\mathrm{old}}_{ik}\) enters both variables with a negative sign. 
Assuming mean-zero rating errors that are independent across models and releases conditional on the true ratings,
\[
\operatorname{Cov} \left( \Delta\hat\theta_{ik},\hat g_{ik} \right) = v^{\mathrm{old}}_{ik}, \qquad \operatorname{Var} \left( \hat g_{ik}-g_{ik} \right) = v^{\mathrm{old}}_{ik}+v^{\mathrm{riv}}_{ik}.
\]
Sampling noise can bias toward a positive catch-up effect. 

We remove this bias using an errors-in-variables correction.
Let \(Z\) denote the design matrix for the developer-category and category-quarter
fixed effects, and let
\[
M = I-Z(Z^\top Z)^{-1}Z^\top
\]
denote the corresponding residual-maker. Thus, multiplying by \(M\) removes
the variation explained by these fixed effects.

After computing residuals, the regression coefficient is the covariance between
the measured competitive gap and subsequent improvement, divided by the
variance of the measured gap. 
Rating noise contributes to both quantities.
We therefore subtract its expected contribution before computing the
coefficient:
\[
\hat\beta_c
=
\frac{
\hat g^\top M\Delta\hat\theta
-
d_{\mathrm{fac}}\,n\bar v^{\mathrm{old}}
}{
\hat g^\top M\hat g
-
d_{\mathrm{fac}}\,n\left(\bar v^{\mathrm{old}}+\bar v^{\mathrm{riv}}\right)
},
\]
where \(\bar v^{\mathrm{old}}\) and \(\bar v^{\mathrm{riv}}\) are the panel-average reported sampling variances for the outgoing model and its rival, respectively, and \(d_{\mathrm{fac}}\) accounts for the degrees of freedom consumed by the within transformation.
The correction has little effect on our estimate:
the baseline estimate is \(\hat\beta=0.487\), compared with
\(\hat\beta_c=0.480\).
This limited change reflects that variation in the measured competitive gaps
is substantially larger than the reported sampling noise.

\subsection{Statistical Comparisons for RQ1}
\label{app:statistical_rq1}
In Figure~\ref{fig:residual_cosine_kde} (left), we compare this distribution to a permutation null that independently permutes developers within each category, preserving each category's marginal distribution of residuals while destroying any developer-specific pattern of relative strength. 
We find little difference between the observed and null distributions: the null has a mean cosine similarity of \(-0.051\), compared with \(-0.048\) in the observed data. 
While this difference is statistically significant (\(p<0.001\), paired \(t\)-test), it is negligible in magnitude.

In Figure~\ref{fig:residual_cosine_kde} (right) we average the similarities between pairs of developers across snapshots. 
Four groupings are particularly visible in the similarity matrix, which we summarize here (each contains additional developers beyond those highlighted): models from IBM and Microsoft are stronger in math, and relatively weaker in multi-turn interactions and creative writing; 
models from NVIDIA and Amazon are stronger in English and coding, and weaker in creative writing; 
models from Google and Meta are stronger in creative writing and multi-turn interactions, and weaker in coding and math;
and models from OpenAI and Anthropic are stronger in instruction following, longer queries, and weaker in English.

\subsection{Statistical Comparisons for RQ2}
\label{app:statistical_rq2}
We define a flagship changeover as any change in the identity of a developer's model with the highest overall rating.
We correct all estimates for measurement error using Arena's reported rating variances, removing the mechanical covariance between the competitive gap and subsequent improvement induced by their shared baseline rating (details in Appendix~\ref{app:noise_correction}).
We restrict this analysis to developers with at least four qualifying flagship changeovers, each with mature ratings (at least 56 days post-release) for the incoming model. 
This leaves 12 developers in our analysis.
Here, $g_{i,k,t}$ is computed as the difference between a model and the best competitor in that category among the 12 developers, at the time that the new model was released (the last snapshot before changing over).

%% file: appendix/ablations.tex
\section{Additional Experiments}
\subsection{Confounds in $\hat{\beta}$}
\label{app:confounds}

\begin{table*}
\caption{We account for confounds when computing $\hat{\beta}$, and find that $\hat{\beta}$ remains positive throughout.}
\label{tab:beta_confounds}
\begin{tabular}{llcrr}
\toprule
& & $\hat\beta$ (s.e.) & Events & $N$ \\
\midrule
\emph{Baseline} & all events & +0.480$^{***}$ (0.117) & 106 & 848 \\
\addlinespace
\emph{Comparator choice} & best rival & +0.480$^{***}$ (0.117) & 106 & 848 \\
& second-best rival & +0.485$^{***}$ (0.120) & 106 & 848 \\
& top-3 average & +0.521$^{***}$ (0.117) & 106 & 848 \\
\addlinespace
\emph{Release year} & 2025 & +0.455$^{***}$ (0.114) & 60 & 480 \\
& 2026 & +0.849$^{***}$ (0.096) & 41 & 328 \\
\addlinespace
\emph{Sample composition} & events A & +0.500$^{***}$ (0.155) & 53 & 424 \\
& events B & +0.331$^{***}$ (0.105) & 53 & 424 \\
\addlinespace
\emph{Event-level heterogeneity} & release-event fixed effects & +0.949$^{***}$ (0.115) & 106 & 848 \\
\midrule
\emph{Regression to the mean} & measured spurious slope & +0.099 (0.038) & 20 splits & 116,257 \\
& correction-implied slope & +0.139 & -- & -- \\
\bottomrule
\end{tabular}
\end{table*}

We next account for other confounds potentially skewing $\hat{\beta}$. 
In Table~\ref{tab:beta_confounds}, we account for confounds including the selection of comparator (e.g., best rival, second-best rival, and top-3 average) and release year (2025 vs. 2026, with 2024 having only $5$ events).
For both of these, we find that $\hat{\beta}>0.45$, showing a similar catch-up effect. 
We additionally test whether the result is driven by a subset of release events by splitting events into two alternating subsamples and computing \(\hat{\beta}\) separately on each. 
We find that \(\hat{\beta}>0.33\) in both halves of the split.
To account for event-specific factors shared across categories, such as differences in the overall magnitude of improvement across releases, we include a release-event fixed effect \(\delta_e\).  
The resulting estimate remains positive and increases to \(\hat\beta=0.949\). 

We finally consider mechanical regression to the mean, since noise in the outgoing model's rating enters both the measured gap and subsequent improvement. 
Because the historical leaderboard provides only rating snapshots, we cannot re-estimate \(\hat{\beta}\) using independent ratings for each release event; instead, we use raw Arena battles to estimate the magnitude of this mechanical effect. 
We retrieve 116,257 battles, randomly split them in half, and fit Bradley--Terry ratings separately on each half to obtain two independent estimates, \(\hat{\theta}^{A}\) and \(\hat{\theta}^{B}\), of each model's rating.
We regress \(\hat{\theta}^{B}-\hat{\theta}^{A}\) on the gap computed from half \(A\) alone. 
Because the two estimates measure the same underlying model performance, any systematic positive relationship should arise from sampling variation in \(\hat\theta^A\), since the two estimates are constructed from independent random halves of the same set of battles.
These battles cover only April--July 2025 and three of our eight categories (plus the overall category), so we use this exercise to estimate the magnitude of the bias rather than to re-estimate \(\hat{\beta}\). 
Averaged over 20 random splits, the spurious slope is \(0.099\), and the errors-in-variables correction predicts $0.139$. 
As a result, measurement error appears too small to mechanically generate the observed catch-up relationship.

\subsection{Decomposing Capability Gaps}
\label{app:strong_competitors}

A larger capability gap can arise either because a developer's outgoing model performs poorly in a capability or because a rival performs well.
To distinguish between these factors, we regress incoming capability on outgoing capability and the rival frontier
separately:
\[
\theta^{\mathrm{new}}_{ikt} = \alpha_{ik}+\lambda_t +\rho\,\theta^{\mathrm{old}}_{ikt} +\delta\,\theta^{\mathrm{rival}}_{ikt} +\epsilon_{ikt},
\]
where $\alpha_{ik}$ are developer-category fixed effects and $\lambda_t$ are quarter fixed effects. 
Unlike our main specification, we use quarter rather than category-quarter fixed effects so that variation in the rival frontier remains identified.

We estimate $\hat\rho=0.363$ and $\hat\delta=-0.078$ (event-clustered SE $=0.116$ for $\hat\delta$). 
Conditional on the outgoing model's capability, we do not find evidence that a stronger rival frontier predicts stronger incoming capability.
Combined with our finding that $\hat{\beta}>0$, our takeaway is that developers improve more in capabilities where they lag, with a larger gap being consistent more strongly with a weaker developer rather than a stronger rival frontier. 
Proposition~\ref{prop:PrimitiveCatchup} provides one mechanism capable of generating the observed catch-up pattern, as diminishing marginal returns makes it plausible for catch-up to occur, although our observational analysis cannot identify the underlying mechanism.
Taking a step back, shared evaluation provides similar incentives to developers, while diminishing returns potentially allows weaker models to catch-up. 

%% file: appendix/equilibrium_proofs.tex
\section{Proofs of Equilibria and Monoculture Results}
\label{app:proofs}
\subsection{Common Cost Function}
\label{app:CommonCost_proof}
\begin{restatable}[Common Cost Functions]{proposition}{CommonCost}
\label{prop:CommonCost}
Suppose each developer $i$ has continuous production function $\Fi$ and cost function $c_i$, where each $c_i:\mathbb{R}_{\ge 0}\to\mathbb{R}_{\ge 0}$ is a strictly increasing bijection with $c_i(0)=0$. Let $c$ be any other such cost function. Then, for each developer $i$, there exists a production function $\tilde{F}_i$ and a bijection between action vectors $\bxi \leftrightarrow\mathbf{z}_i$ such that

$$
c_i\big(\lVert\mathbf{x}_i\rVert_1\big)=c\big(\lVert\mathbf{z}_i\rVert_1\big)
\qquad\text{and}\qquad
\Fi(\mathbf{x}_i)=\tilde{F}_i(\mathbf{z}_i).
$$

Consequently, the game defined by ${(\Fi,c_i)}_{i=1}^{N}$ is equivalent under reparameterization to the game defined by ${(\tilde \Fi,c)}_{i=1}^{N}$: corresponding strategy profiles induce identical payoffs, and their Nash equilibria are in bijective correspondence.
\end{restatable}
\begin{proof}
Define the scalar map

$$
g_i(T) := c_i^{-1}\big(c(T)\big), \qquad T\ge 0.
$$

Since $c_i$ and $c$ are strictly increasing bijections of $\mathbb{R}_{\ge 0}$, so is $g_i$. In particular, $g_i^{-1}$ exists and is itself a strictly increasing bijection of $\mathbb{R}_{\ge 0}$. By construction,
\begin{equation}
c_i\big(g_i(T)\big) = c(T) \qquad \text{for all } T\ge 0. \label{eq:def_c_i}
\end{equation}
Equivalently, substituting $g_i^{-1}(T)$ for $T$ gives
\begin{equation}
c\big(g_i^{-1}(T)\big) = c_i(T) \qquad \text{for all } T\ge 0. \label{eq:def_c}
\end{equation}

For $\mathbf{x}_i \neq \mathbf{0}$ define the \emph{reference-unit} vector

$$
\mathbf{z}_i := \frac{g_i^{-1}\big(\lVert \mathbf{x}_i\rVert_1\big)}{\lVert \mathbf{x}_i\rVert_1}\,\mathbf{x}_i,
$$

and set $\mathbf{z}_i=\mathbf{0}$ when $\mathbf{x}_i=\mathbf{0}$. Thus $\mathbf{x}_i$ is rescaled, keeping its direction fixed, so that $\lVert \mathbf{z}_i\rVert_1 = g_i^{-1}(\lVert \mathbf{x}_i\rVert_1)$. Applying $g_i$ to both sides of this identity gives

$$
g_i\big(\lVert \mathbf{z}_i\rVert_1\big) = \lVert \mathbf{x}_i\rVert_1.
$$

Moreover, by \eqref{eq:def_c},

$$
c\big(\lVert \mathbf{z}_i\rVert_1\big) = c\big(g_i^{-1}(\lVert \mathbf{x}_i\rVert_1)\big) = c_i\big(\lVert \mathbf{x}_i\rVert_1\big),
$$

so cost is exactly preserved.

For $\mathbf{z}\neq\mathbf{0}$, define

$$
\tilde{F}_i(\mathbf{z}) := F_i\!\left(\frac{g_i\big(\lVert \mathbf{z}\rVert_1\big)}{\lVert \mathbf{z}\rVert_1}\,\mathbf{z}\right),
$$

and let $\tilde{F}_{i}(\mathbf{0}) = F_{i}(\mathbf{0})$. 
Evaluating at $\mathbf{z}=\mathbf{z}_i$ for $\mathbf{x}_{i} \neq \mathbf{0}$ using $g_i\big(\lVert \mathbf{z}_i\rVert_1\big) = \lVert \mathbf{x}_i\rVert_1
$ gives,

$$
\tilde{F}_i(\mathbf{z}_i) = F_i\!\left(\frac{\lVert \mathbf{x}_i\rVert_1}{\lVert \mathbf{z}_i\rVert_1}\cdot\frac{g_i^{-1}(\lVert \mathbf{x}_i\rVert_1)}{\lVert \mathbf{x}_i\rVert_1}\,\mathbf{x}_i\right) = F_i(\mathbf{x}_i),
$$

where we used $\lVert \mathbf{z}_i\rVert_1 = g_i^{-1}(\lVert \mathbf{x}_i\rVert_1)$. If $\mathbf{x}_i=\mathbf{0}$, then $\mathbf{z}_i=\mathbf{0}$ and the equality follows directly from $\tilde F_i(\mathbf{0})=F_i(\mathbf{0})$. Thus $F_i(\mathbf{x}_i)=\tilde F_i(\mathbf{z}_i)$ as required.

Finally, the map $\mathbf{x}_i\mapsto\mathbf{z}_i$ is a bijection of $\mathbb{R}_{\ge0}^K$, with inverse

$$
\mathbf{x}_i = \frac{g_i(\lVert \mathbf{z}_i\rVert_1)}{\lVert \mathbf{z}_i\rVert_1}\mathbf{z}_i
$$

for $\mathbf{z}_i\neq\mathbf{0}$, and $\mathbf{x}_i=\mathbf{0}$ when $\mathbf{z}_i=\mathbf{0}$.

Applying this construction to every developer $i$, corresponding strategy profiles $(\mathbf{x}_1,\ldots,\mathbf{x}_N)$ and $(\mathbf{z}_1,\ldots,\mathbf{z}_N)$ induce identical outputs and costs for every developer, and hence identical payoffs. Since the correspondence is a bijection for each developer, best responses are preserved under the reparameterization, and therefore the Nash equilibria of the two games are in bijective correspondence.
\end{proof}

\subsection{Proof of Theorem~\ref{thm:EquilibriumExistence}}
\label{app:EquilibriumExistence_proof}
\EquilibriumExistence*
\begin{proof}
Since $\sigma$ is bounded in $[0,1]$ and $\mathbf{w}\in\Delta^{K-1}$, the total reward of any developer is at most $1$. Moreover, choosing $\mathbf{x}_i=\mathbf{0}$ gives nonnegative utility. Therefore, any action satisfying
\[
c(\lVert\mathbf{x}_i\rVert_1)>1
\]
cannot be a best response. Let $\bar{x}=c^{-1}(1)$. We may therefore restrict each developer's strategy set to
\[
\mathcal{X}_i=\{\mathbf{x}_i\in\mathbb{R}_+^K:\lVert\mathbf{x}_i\rVert_1\leq\bar{x}\},
\]
which is compact and convex.
Since any action outside \(\mathcal X_i\) yields strictly negative utility, while \(\mathbf x_i=\mathbf0\) yields nonnegative utility, no action outside \(\mathcal X_i\) can be a profitable deviation. Thus any Nash equilibrium of the restricted game is also a Nash equilibrium of the original game.
We note that $u_{i}$ is continuous because it is built from the addition and multiplication of continuous functions in $\sigma$, $F_{i}$, and $c$. 

It remains to show that $u_i(\mathbf{x}_i,\mathbf{x}_{-i})$ is  concave in $\mathbf{x}_i$. For $z_{ij,k}=g_{i,k}(\mathbf{x}_i)-g_{j,k}(\mathbf{x}_j)$,
\[
\nabla_{\mathbf{x}_i}^2 \sigma(z_{ij,k})
=
\sigma'(z_{ij,k})\nabla^2 g_{i,k}(\mathbf{x}_i)
+
\sigma''(z_{ij,k})
\nabla g_{i,k}(\mathbf{x}_i)\nabla g_{i,k}(\mathbf{x}_i)^\top.
\]
Since $\sigma'(z)>0$ and
\[
\frac{\sigma''(z)}{\sigma'(z)} \leq \rho_\sigma,
\]
the assumed curvature condition implies
\[
\nabla_{\mathbf{x}_i}^2 \sigma(z_{ij,k})\preceq 0.
\]
Thus every pairwise reward term is concave in $\mathbf{x}_i$. Their nonnegative weighted sum is therefore concave, and since $c$ is convex,
\[
-c(\lVert\mathbf{x}_i\rVert_1)
\]
is concave as well. Hence $u_i(\cdot,\mathbf{x}_{-i})$ is concave (and so quasi-concave) on $\mathcal{X}_i$ for every $\mathbf{x}_{-i}$.

Finally, the utilities are continuous and the strategy sets $\mathcal{X}_i$ are nonempty, compact, and convex. Therefore, by the Debreu--Fan--Glicksberg equilibrium existence theorem, the game admits a pure-strategy Nash equilibrium.
\end{proof}

\subsection{Proof of Lemma~\ref{lem:SymmetricDevelopers}}
\label{app:SymmetricDevelopers_proof}
\SymmetricDevelopers*
\begin{proof}
By Theorem~\ref{thm:EquilibriumExistence}, we may restrict each developer's action space to a compact convex set $\mathcal{X}_i$, and each payoff $u_i$ is continuous and concave in the developer's own action. Hence each best-response correspondence is nonempty, convex-valued, and upper hemicontinuous.

First suppose that $F_i=F_{i'}$. Consider the subset of strategy profiles
\[
\mathcal{X}^{i,i'}
=
\left\{
\bx \in\prod_{j=1}^N\mathcal{X}_j:
\bxi =\bx_{i'}
\right\}.
\]
At any profile $\bx \in\mathcal{X}^{i,i'}$, developers $i$ and $i'$ face identical optimization problems: they have the same production function and cost function, and, because $\bxi=\bx_{i'}$, each faces the same actions and outputs from all other developers. Therefore,
\[
\BR_i(\bxmi)
=
\BR_{i'}(\bx_{-i'}).
\]
Since these best-response sets are identical, nonempty, and convex, we can define a reduced best-response correspondence on $\mathcal{X}^{i,i'}$ that requires developers $i$ and $i'$ to select the same element of their common best-response set, while every other developer selects from its usual best-response set. 
Equivalently, parameterize $\mathcal{X}^{i,i'}$ by a common action $\mathbf y=\mathbf x_i=\mathbf x_{i'}$ and the actions $(\mathbf x_\ell)_{\ell\neq i,i'}$. 
The resulting correspondence is nonempty and convex-valued. 
Moreover, it is upper hemicontinuous because each developer's best-response correspondence is upper hemicontinuous and the restriction to profiles satisfying $\mathbf x_i=\mathbf x_{i'}$ is continuous.
Thus, by Kakutani's fixed-point theorem, there exists a fixed point $\bx^*\in\mathcal{X}^{i,i'}$.
At this fixed point every developer is playing a best response in the original game, and, by construction, $\bxi^*=\bx_{i'}^*$.

Now suppose that $F_i=F$ for every $i$. 
Now consider the fully symmetric subset
\[
\mathcal{X}^{\mathrm{sym}}
=
\left\{
\bx\in\prod_{j=1}^N\mathcal{X}_j:
\bx_1=\cdots=\bx_N
\right\}.
\]
Here, every developer faces the same optimization problem, and therefore faces the same best response problem. 
We can choose the same best response for each developer under this situation. 
As above, the resulting correspondence maps $\mathcal{X}^{\mathrm{sym}}$ into itself and satisfies the conditions of Kakutani's fixed-point theorem.
Therefore, there exists a fixed point with $\bxi$ constant, which implies $\bthetai$ constant, and consequently $H=1$.
\end{proof}

\subsection{Proof of Theorem~\ref{thm:SimilarFHomogeneity}}
\label{app:SimilarFHomogeneity}
\SimilarFHomogeneity*
\begin{proof}
Let $G(\bx)$ be the pseudogradient for the original game with production functions $F_{i}$ and $F_{i'}$ and let $\widetilde{G}(\bx)$ be the pseudogradient for the counterfactual game where $F_{i}$ is replaced with $F_{i'}$ (while everything else stays constant). 

We show that $G$ and $\widetilde{G}$ are close in L2 norm. 
For each developer $\ell$, the corresponding block of the pseudogradient is
\[
G_\ell(\bx)=-\nabla_{\bx_\ell}u_\ell(\bx).
\]
Since
\[
u_\ell(\bx)=\frac{1}{N-1}\sum_{r\neq \ell}\sum_{k=1}^K w_k\sigma\!\left(F_\ell(\bx_\ell)_k-F_r(\bx_r)_k\right)-c(\lVert\bx_\ell\rVert_1),
\]
we have
\[
\nabla_{\bx_\ell}u_\ell(\bx)=\frac{1}{N-1}\sum_{r\neq \ell}\sum_{k=1}^K w_k \sigma' \!\left(F_\ell(\bx_\ell)_k-F_r(\bx_r)_k\right)\nabla_{\bx_\ell}F_\ell(\bx_\ell)_k-\nabla_{\bx_\ell}c(\lVert\bx_\ell\rVert_1).
\]
Since the strategy space is compact and the production functions are continuous, the arguments of $\sigma$ lie in a compact interval.
Since $\sigma$ is twice continuously differentiable, $\sigma'$ and $\sigma''$ are bounded on this interval, so $\sigma'$ is Lipschitz; moreover, since the production functions are continuously differentiable on the compact strategy space, their gradients are uniformly bounded.
Replacing $F_i$ with $F_{i'}$ affects only terms involving developer $i$.
For $\ell=i$, both the argument of $\sigma'$ and the production-function gradient change; for $\ell\neq i$, only the comparison against developer $i$ changes. 
Using $\lVert F_i(\bx_i)-F_{i'}(\bx_i)\rVert_2\leq\epsilon$, $\lVert \nabla F_i(\bx_i)_k-\nabla F_{i'}(\bx_i)_k\rVert_2\leq\epsilon$, and the Lipschitz continuity of $\sigma'$, we can bound these differences term-by-term to obtain
\[
\sup_{\bx\in\mathcal X}\lVert G(\bx)-\widetilde G(\bx)\rVert_2 \leq C\epsilon,
\] 
for some constant $C>0$ independent of $\epsilon$.

Next, let $\bx^*$ denote the equilibrium of the original game. 
Because $F_i$ and $F_{i'}$ both satisfy the conditions of Theorem~\ref{thm:EquilibriumExistence}, the counterfactual game also satisfies these conditions. 
By Lemma~\ref{lem:SymmetricDevelopers}, there exists an equilibrium $\widetilde{\bx}^*$ of the counterfactual game such that $\widetilde{\bx}_i^*=\widetilde{\bx}_{i'}^*$. 
We select such an equilibrium $\widetilde{\bx}^*$.
Since each developer's utility is concave and differentiable in its own
strategy, the first-order optimality conditions characterize equilibrium.
That is, because $\bx^*$ and $\widetilde{\bx}^*$ are equilibria of the original and counterfactual games, respectively, we have that 
\[
G(\bx^*)^\top(\widetilde{\bx}^*-\bx^*)\geq0,\qquad \widetilde G(\widetilde{\bx}^*)^\top(\bx^*-\widetilde{\bx}^*)\geq0.
\]
Let $d=\bx^*-\widetilde{\bx}^*$. Adding the two inequalities gives $(G(\bx^*)-\widetilde G(\widetilde{\bx}^*))^\top d\leq0$. 
Recall that $G(\mathbf{x})$ is strongly monotone, and so 
\[
\left(G(\bx)-G(\bx')\right)^\top(\bx-\bx')
\geq
\mu\lVert\bx-\bx'\rVert_2^2.
\]
Setting $\bx=\bx^*$ and $\bx'=\widetilde{\bx}^*$ gives
\[
(G(\bx^*)-G(\widetilde{\bx}^*))^\top d\geq\mu\lVert d\rVert_2^2.
\]
Invoking $(G(\bx^*)-\widetilde G(\widetilde{\bx}^*))^\top d\leq0$ gives that 
\[
\mu\lVert d\rVert_2^2\leq(G(\bx^*)-G(\widetilde{\bx}^*))^\top d\leq(\widetilde G(\widetilde{\bx}^*)-G(\widetilde{\bx}^*))^\top d \leq C\epsilon \lVert d\rVert_2,
\]
where the final inequality follows from Cauchy--Schwarz and the bound on the difference between the pseudogradients.
Rearranging and solving for $d$ gives that 
\[
\lVert\bx^*-\widetilde{\bx}^*\rVert_2\leq\frac{C\epsilon}{\mu}.
\]

Since $\widetilde{\bx}_i^*=\widetilde{\bx}_{i'}^*$,
\[
\lVert\bx_i^*-\bx_{i'}^*\rVert_2\leq \lVert\bx_i^*-\widetilde{\bx}_i^*\rVert_2+\lVert\bx_{i'}^*-\widetilde{\bx}_{i'}^*\rVert_2\leq O(\lVert\bx^*-\widetilde{\bx}^*\rVert_2) =O\left(\frac{\epsilon}{\mu}\right).
\]

Finally, to bound $\Hij(\mathbf{x}^*)$, let 
\[
\begin{aligned}
\delta
&=
\lVert\btheta_i^*-\btheta_{i'}^*\rVert_2\\
&=
\lVert F_i(\bx_i^*)-F_{i'}(\bx_{i'}^*)\rVert_2\\
&\leq
\lVert F_i(\bx_i^*)-F_i(\bx_{i'}^*)\rVert_2
+
\lVert F_i(\bx_{i'}^*)-F_{i'}(\bx_{i'}^*)\rVert_2\\
&\leq
L\lVert\bx_i^*-\bx_{i'}^*\rVert_2+\epsilon\\
&=
O\left(\epsilon\left(1+\frac{L}{\mu}\right)\right).
\end{aligned}
\]
Next, recall that
\[
H_{i,i'}(\mathbf{x}^*)=\frac{\langle\btheta_i^*,\btheta_{i'}^*\rangle}{\lVert\btheta_i^*\rVert_2\lVert\btheta_{i'}^*\rVert_2}.
\]
Let $\widehat{\btheta}_i^*=\btheta_i^*/\lVert\btheta_i^*\rVert_2$ and $\widehat{\btheta}_{i'}^*=\btheta_{i'}^*/\lVert\btheta_{i'}^*\rVert_2$. Since both normalized vectors have unit norm,
\[
\lVert\widehat{\btheta}_i^*-\widehat{\btheta}_{i'}^*\rVert_2^2
=2-2\langle\widehat{\btheta}_i^*,\widehat{\btheta}_{i'}^*\rangle
=2(1-H_{i,i'}(\mathbf{x}^*)).
\]
If $\lVert\btheta_i^*\rVert_2,\lVert\btheta_{i'}^*\rVert_2\geq m$, then, adding and subtracting $\btheta_{i'}^*/\lVert\btheta_i^*\rVert_2$ and applying the triangle inequality,
\[
\begin{aligned}
\left\lVert\frac{\btheta_i^*}{\lVert\btheta_i^*\rVert_2}-\frac{\btheta_{i'}^*}{\lVert\btheta_{i'}^*\rVert_2}\right\rVert_2
&=
\left\lVert
\frac{\btheta_i^*-\btheta_{i'}^*}{\lVert\btheta_i^*\rVert_2}
+
\btheta_{i'}^*
\left(
\frac{1}{\lVert\btheta_i^*\rVert_2}
-
\frac{1}{\lVert\btheta_{i'}^*\rVert_2}
\right)
\right\rVert_2 \\
&\leq
\frac{\lVert\btheta_i^*-\btheta_{i'}^*\rVert_2}{\lVert\btheta_i^*\rVert_2}
+
\lVert\btheta_{i'}^*\rVert_2
\left|
\frac{1}{\lVert\btheta_i^*\rVert_2}
-
\frac{1}{\lVert\btheta_{i'}^*\rVert_2}
\right| \\
&=
\frac{\lVert\btheta_i^*-\btheta_{i'}^*\rVert_2}{\lVert\btheta_i^*\rVert_2}
+
\frac{\left|\lVert\btheta_i^*\rVert_2-\lVert\btheta_{i'}^*\rVert_2\right|}{\lVert\btheta_i^*\rVert_2} \\
&\leq
\frac{2\lVert\btheta_i^*-\btheta_{i'}^*\rVert_2}{m}
=
\frac{2\delta}{m}.
\end{aligned}
\]
Therefore,
\[
2(1-H_{i,i'}(\mathbf{x}^*))\leq\frac{4\delta^2}{m^2},
\]
and hence
\[
H_{i,i'}(\mathbf{x}^*)\geq1-\frac{2\delta^2}{m^2}=1-O\left(\frac{\epsilon^2}{m^2}\left(1+\frac{L}{\mu}\right)^2\right).
\]
\end{proof}

\subsection{Proof of Theorem~\ref{thm:SpecializedFDifferentiation}}
\label{app:SpecializedFDifferentiation}
\SpecializedFDifferentiation*

\begin{proof}
Let
\[
    Y_i^*
    :=
    \{\theta_j^*:j\neq i\},
    \qquad
    Y_{i'}^*
    :=
    \{\theta_j^*:j\neq i'\}
\]
denote the rival environments faced by developers $i$ and $i'$ at
equilibrium. Since $\mathbf x^*$ is a Nash equilibrium and $\Theta$
contains all relevant best responses, we have
\[
    \theta_i^*
    =
    \arg\min_{\theta\in\Theta}
    \phi_{i,Y_i^*}(\theta),
    \qquad
    \theta_{i'}^*
    =
    \arg\min_{\theta\in\Theta}
    \phi_{i',Y_{i'}^*}(\theta).
\]
By the assumed interiority of the minimizers,
\[
    \nabla \phi_{i,Y_i^*}(\theta_i^*)=0,
    \qquad
    \nabla \phi_{i',Y_{i'}^*}(\theta_{i'}^*)=0.
\]

We first compare the two developers while holding the rival environment
fixed. Define the counterfactual best response of developer $i'$ to
developer $i$'s rival environment:
\[
    \widetilde\theta_{i'}
    :=
    \arg\min_{\theta\in\Theta}
    \phi_{i',Y_i^*}(\theta).
\]
We will decompose $\lVert \theta^*_i-\theta^*_{i'} \rVert_2$ into $\lVert \theta^*_i-\widetilde{\theta}_{i'} \rVert_2$ and $\lVert \widetilde{\theta}_{i'} -\theta^*_{i'}\rVert_2$.

To bound the first term, we note that $\nabla\phi_{i,Y_i^*}(\theta_i^*)=0$ implies $\nabla C_{i}(\theta_{i}^*) = \nabla V_{Y_{i}^*}(\theta_{i}^*)$. 
As a result, 
\[
\begin{aligned}
    \lVert \nabla\phi_{i',Y_i^*}(\theta_i^*) \rVert_2
    &=
    \lVert\nabla C_{i'}(\theta_i^*)
    -
    \nabla V_{Y_i^*}(\theta_i^*) \rVert_2\\
    &=
    \lVert \nabla C_{i'}(\theta_i^*)
    -
    \nabla C_i(\theta_i^*) \rVert_2 \\
    & \geq \delta_{i,i'} 
\end{aligned}
\]
We additionally note that by interiority, $\nabla\phi_{i',Y_i^*}(\widetilde\theta_{i'})=0$. 
Therefore, combining these two gives that 
\[
 \delta_{i,i'}
    \le
    \left\|
    \nabla\phi_{i',Y_i^*}(\theta_i^*)
    -
    \nabla\phi_{i',Y_i^*}(\widetilde\theta_{i'})
    \right\|_2.
\]
We can upper bound this because $\phi_{i',Y_{i}^*}$ has L-Lipschitz gradient, and so 
\[
 \delta_{i,i'}
    \le
    \left\|
    \nabla\phi_{i',Y_i^*}(\theta_i^*)
    -
    \nabla\phi_{i',Y_i^*}(\widetilde\theta_{i'})
    \right\|_2 \leq  L\|\theta_i^*-\widetilde\theta_{i'}\|_2.
\]
Rearranging gives,
\[
    \|\theta_i^*-\widetilde\theta_{i'}\|_2
    \ge
    \frac{\delta_{i,i'}}{L}.
\]

Next, we need to bound $\lVert\widetilde\theta_{i'}-\theta_{i'}^* \rVert_2$. 
The environments $Y_i^*$ and $Y_{i'}^*$ share the same $N-2$ other rivals and differ only in that $Y_i^*$ contains $\theta_{i'}^*$ whereas $Y_{i'}^*$ contains $\theta_i^*$. 
Thus, for any $\theta\in\Theta$ and coordinate $k$,
\[
\begin{aligned}
&
\left|
\frac{\partial V_{Y_i^*}(\theta)}{\partial\theta_k}
-
\frac{\partial V_{Y_{i'}^*}(\theta)}{\partial\theta_k}
\right| \\
&\qquad =
\frac{w_k}{N-1}
\left|
\sigma'(\theta_k-\theta_{i',k}^*)
-
\sigma'(\theta_k-\theta_{i,k}^*)
\right| \leq \frac{\ell_\sigma w_k}{N-1}
|\theta_{i,k}^*-\theta_{i',k}^*|.
\end{aligned}
\]
Summing across coordinates gives
\[
\begin{aligned}
\left\|
\nabla V_{Y_i^*}(\theta)
-\nabla V_{Y_{i'}^*}(\theta)
\right\|_2^2
&\le
\frac{\ell_\sigma^2}{(N-1)^2}
\sum_{k=1}^K
w_k^2
|\theta_{i,k}^*-\theta_{i',k}^*|^2\\
&\le
\frac{\ell_\sigma^2\|w\|_\infty^2}{(N-1)^2}
\|\theta_i^*-\theta_{i'}^*\|_2^2,
\end{aligned}
\]
and so
\[
    \left\|
    \nabla V_{Y_i^*}(\theta)
    -
    \nabla V_{Y_{i'}^*}(\theta)
    \right\|_2
    \le
    \frac{\ell_\sigma\|w\|_\infty}{N-1}
    \|\theta_i^*-\theta_{i'}^*\|_2.
\]

Because the cost term is the same for both $\phi_{i',Y_{i}^*}$ and $\phi_{i',Y_{i'}^*}$, we have that 
\[
\begin{aligned}
    \left\|
    \nabla\phi_{i',Y_i^*}(\theta_{i'}^*)
    \right\|_2
    &=
    \left\|
    \nabla\phi_{i',Y_i^*}(\theta_{i'}^*)
    -
    \nabla\phi_{i',Y_{i'}^*}(\theta_{i'}^*)
    \right\|_2 \\
    &=
    \left\|
    \nabla V_{Y_{i'}^*}(\theta_{i'}^*)
    -
    \nabla V_{Y_i^*}(\theta_{i'}^*)
    \right\|_2 \\
    &\le
    \frac{\ell_\sigma\|w\|_\infty}{N-1}
    \|\theta_i^*-\theta_{i'}^*\|_2.
\end{aligned}
\]

Since $\phi_{i',Y_i^*}$ is $\mu$-strongly convex and
$\widetilde\theta_{i'}$ is its minimizer, strong monotonicity of its
gradient gives
\[
\begin{aligned}
    \mu
    \|\theta_{i'}^*-\widetilde\theta_{i'}\|_2^2
    &\le
    \left\langle
    \nabla\phi_{i',Y_i^*}(\theta_{i'}^*)
    -
    \nabla\phi_{i',Y_i^*}(\widetilde\theta_{i'}),
    \theta_{i'}^*-\widetilde\theta_{i'}
    \right\rangle \\
    &\le
    \left\|
    \nabla\phi_{i',Y_i^*}(\theta_{i'}^*)
    \right\|_2
    \|\theta_{i'}^*-\widetilde\theta_{i'}\|_2.
\end{aligned}
\]
Therefore,
\[
    \|\theta_{i'}^*-\widetilde\theta_{i'}\|_2
    \le
    \frac{1}{\mu}
    \left\|
    \nabla\phi_{i',Y_i^*}(\theta_{i'}^*)
    \right\|_2 \leq \frac{1}{\mu} \frac{\ell_\sigma\|w\|_\infty}{N-1}
    \|\theta_i^*-\theta_{i'}^*\|_2 = \rho_{N} \lVert \theta_i^*-\theta_{i'}^* \rVert_2,
\]
where
\[
    \rho_N
    =
    \frac{\ell_\sigma\|w\|_\infty}
         {\mu(N-1)}.
\]

Finally, by the triangle inequality,
\[
\begin{aligned}
    \|\theta_i^*-\widetilde\theta_{i'}\|_2
    &\le
    \|\theta_i^*-\theta_{i'}^*\|_2
    +
    \|\theta_{i'}^*-\widetilde\theta_{i'}\|_2 \\
    &\le
    (1+\rho_N)
    \|\theta_i^*-\theta_{i'}^*\|_2.
\end{aligned}
\]
Plugging in the bound for $\lVert \theta^*_i-\widetilde{\theta}_{i'} \rVert_2$ yields
\[
    \|\theta_i^*-\theta_{i'}^*\|_2
    \ge
    \frac{\delta_{i,i'}}
         {L(1+\rho_N)}.
\]

For the cosine-similarity bound, let
\[
    d:=\|\theta_i^*-\theta_{i'}^*\|_2,
    \qquad
    r_i:=\|\theta_i^*\|_2,
    \qquad
    r_{i'}:=\|\theta_{i'}^*\|_2.
\]
For nonzero outputs,
\[
\begin{aligned}
    d^2
    &=
    r_i^2+r_{i'}^2
    -
    2r_ir_{i'}H_{i,i'}(\mathbf x^*)\\
    &=
    (r_i-r_{i'})^2
    +
    2r_ir_{i'}
    \bigl(1-H_{i,i'}(\mathbf x^*)\bigr).
\end{aligned}
\]
Therefore,
\[
    H_{i,i'}(\mathbf x^*)
    =
    1-
    \frac{
        d^2-(r_i-r_{i'})^2
    }{
        2r_ir_{i'}
    }.
\]
Using
\[
    d
    \ge
    \frac{\delta_{i,i'}}
         {L(1+\rho_N)}
\]
and the trivial bound $H_{i,i'}\le1$ gives
\[
H_{i,i'}(\mathbf x^*)
\le
1-
\frac{
\left[
\left(
\frac{\delta_{i,i'}}{L(1+\rho_N)}
\right)^2
-(r_i-r_{i'})^2
\right]_+
}{
2r_ir_{i'}
}.
\]
\end{proof}

\subsection{Proof of Proposition~\ref{prop:DirectionalCharacterization}}
\label{app:DirectionalCharacterization}
\DirectionalCharacterization*
\begin{proof}
First suppose that $D=(\mathbf d_1,\ldots,\mathbf d_N)$ is induced by a pure-strategy Nash equilibrium $\bx^*$ with nonzero outputs.
Let $\alpha_i=\lVert F_i(\bx_i^*)\rVert_2$ and  $\mathbf d_i = \frac{F_i(\bx_i^*)}{\lVert F_i(\bx_i^*)\rVert_2}$, so that $F_i(\bx_i^*)=\alpha_i\mathbf d_i$
for every developer $i$.

For each developer $i$ and capability $k$, define its marginal evaluation weight at
$\bx^*$ as
\[
p_{i,k}^*
=
\frac{w_k}{N-1}
\sum_{j\neq i}
\sigma'\!\left(
F_i(\bx_i^*)_k-F_j(\bx_j^*)_k
\right).
\]
Since $F_i(\bx_i^*)=\alpha_i\mathbf d_i$, this is exactly
\[
p_{i,k}^*
=
p_{i,k}(D,\boldsymbol{\alpha})
=
\frac{w_k}{N-1}
\sum_{j\neq i}
\sigma'\!\left(
\alpha_i d_{i,k}-\alpha_j d_{j,k}
\right).
\]

The gradient of developer $i$'s utility with respect to its investment is therefore
\[
\nabla_{\bx_i}u_i(\bx^*)
=
\nabla F_i(\bx_i^*)^\top
\mathbf p_i(D,\boldsymbol{\alpha})
-
c'(\lVert\bx_i^*\rVert_1)\mathbf 1.
\]
Because $\bx^*$ is an equilibrium, the first-order optimality conditions imply
\[
\nabla F_i(\bx_i^*)^\top
\mathbf p_i(D,\boldsymbol{\alpha})
\leq
c'(\lVert\bx_i^*\rVert_1)\mathbf 1,
\]
with equality in every coordinate $\ell$ such that $x_{i,\ell}^*>0$.
Thus $D$ is locally supportable.

Conversely, suppose that $D$ is locally supportable.
Then there exist $\boldsymbol{\alpha}\in\mathbb R_{++}^N$ and
$\bx=(\bx_1,\ldots,\bx_N)$ such that
\[
F_i(\bx_i)=\alpha_i\mathbf d_i
\]
and
\[
\nabla F_i(\bx_i)^\top
\mathbf p_i(D,\boldsymbol{\alpha})
\leq
c'(\lVert\bx_i\rVert_1)\mathbf 1,
\]
with equality whenever $x_{i,\ell}>0$.

At this profile,
\[
p_{i,k}(D,\boldsymbol{\alpha})
=
\frac{w_k}{N-1}
\sum_{j\neq i}
\sigma'\!\left(
F_i(\bx_i)_k-F_j(\bx_j)_k
\right),
\]
and hence the conditions above are exactly the first-order optimality
conditions for developer $i$'s original payoff $u_i(\bx_i,\bx_{-i})$.

By the curvature argument in the proof of Theorem~\ref{thm:EquilibriumExistence}, \(u_i(\cdot,\bx_{-i})\) is concave on \(\mathbb R_+^K\). 
Therefore, these first-order conditions are sufficient for optimality so $\bx_i$ is a best response to $\bx_{-i}$ for every developer $i$.
It follows that $\bx$ is a pure-strategy Nash equilibrium, and its normalized
output directions are exactly $D$.

Finally, complete monoculture occurs if and only if
\[
\mathbf d_1=\cdots=\mathbf d_N,
\]
so the first characterization follows by setting
$D=(\mathbf d,\ldots,\mathbf d)$.
Similarly, because output vectors are nonnegative, complete specialization
occurs if and only if
\[
\mathbf d_i^\top\mathbf d_j=0
\qquad\forall i\neq j.
\]
The second characterization therefore follows directly.
\end{proof}

\subsection{Proof of Proposition~\ref{prop:AsymptoticBunching}}
\label{app:AsymptoticBunching}
\AsymptoticBunching*

\begin{proof}
Recall that at each time $t$, developer $i$ chooses $\mathbf x_{i,t}$ to maximize the win-rate, minus an unbounded cost function.
The win-rate is at most $1$, while the cost function is increasing and unbounded, so there exists $B<\infty$ such that $c(B)>1$. 
Therefore, at any equilibrium, $\|\mathbf x_{i,t}\|_1\le B$ for every $i,t$. 

Next, we deal with two cases on whether the cumulative investments $\{\mathbf X_i^t\}_{t\ge1}$ are bounded or not, and show that in either case $\|F_i(\mathbf X_i^{t+1})-F_i(\mathbf X_i^t)\|_2 \to 0$.  
First, suppose that $\{\mathbf X_i^t\}_{t\ge1}$ is bounded. 
Because $\mathbf X_i^t$ is the sum of $t$ non-negative numbers, if sum converges, then the individual terms must go to zero; $\sum_t\|\mathbf x_{i,t}\|_1<\infty$, and so $\|\mathbf x_{i,t}\|_1\to0$.
Because $\|\mathbf x_{i,t}\|_1\to0$, then $\lVert \mathbf X_i^{t+1}-\mathbf X_i^t \rVert_1 \to 0$. 
Because $F_i$ is continuously differentiable, its Jacobian is bounded on the compact set containing $\{\mathbf X_i^t\}_{t\ge1}$, and so 
\[
\|F_i(\mathbf X_i^{t+1})-F_i(\mathbf X_i^t)\|_2
\to0.
\]

Instead, suppose that $\{\mathbf X_i^t\}_{t\ge1}$ is unbounded, so  $\|\mathbf X_i^t\|_1\to\infty$.
We know that $\lVert X_i^{t+1}-X_i^{t} \rVert_1 \leq B$.
We can then use the mean-value inequality to show that 
\[
\|F_i(\mathbf X_i^{t+1})-F_i(\mathbf X_i^t)\|_2
\le
B
\sup_{\|\mathbf z\|_1\ge\|\mathbf X_i^t\|_1}
\max_{\ell\in[K]}
\left\|
\frac{\partial F_i(\mathbf z)}{\partial z_\ell}
\right\|_2 .
\]
As $t\to\infty$, $\|\mathbf X_i^t\|_1\to\infty$, and by assumption, we have that $\left\|
\frac{\partial F_i(\mathbf z)}{\partial z_\ell}
\right\|_2 \to 0$.
As a result, $\|F_i(\mathbf X_i^{t+1})-F_i(\mathbf X_i^t)\|_2 \to 0$, so $\|\Delta\boldsymbol{\theta}^t\|_2\to0$.

We next use the catch-up property. 
By Cauchy--Schwarz, we have that 
\[
\beta\|\mathbf g^t\|_2^2
\le
\left\langle
\mathbf g^t,\Delta\boldsymbol{\theta}^t
\right\rangle
\le
\|\mathbf g^t\|_2
\|\Delta\boldsymbol{\theta}^t\|_2.
\]
If $\|\mathbf g^t\|_2>0$, dividing by $\|\mathbf g^t\|_2$ gives
\[
\|\mathbf g^t\|_2
\le
\frac{1}{\beta}
\|\Delta\boldsymbol{\theta}^t\|_2.
\]
Because $\|\Delta\boldsymbol{\theta}^t\|_2 \to 0$ and $\beta$ is a constant, we have that $\lVert \mathbf{g}^t \rVert_2\to 0$. 
By definition of $\mathbf{g}$ this implies that 
\[
g_{i,k}^t
=
\max_{i'}\theta_{i',k}^t-\theta_{i,k}^t
\to0
\qquad\text{for every }i,k.
\]
\end{proof}

\subsection{Proof of Proposition~\ref{prop:PrimitiveCatchup}}
\label{app:PrimitiveCatchup}
\PrimitiveCatchup*
\begin{proof}
Fix a developer $i$, category $k$, and period $t$.  Write
\[
    \theta_{r,k}^t := [F_r(\mathbf X_r^t)]_k,
    \qquad
    \Delta\theta_{r,k}^t
    :=
    \theta_{r,k}^{t+1}-\theta_{r,k}^t,
\]
and let $j\in\arg\max_r \theta_{r,k}^t$ be a developer attaining
the category-$k$ frontier at time $t$. 
That is, $g_{i,k}^t=\theta_{j,k}^t-\theta_{i,k}^t$.
If $g_{i,k}^t=0$, the result follows immediately from coordinatewise
monotonicity of $F_i$, since $\Delta\theta_{i,k}^t\ge 0$.  

Next, suppose $g_{i,k}^t>0$.
We first lower bound the marginal evaluation weight on every category.
For any developers $r,s$, category $q$, and relevant states
$\mathbf x,\mathbf y$, suppose without loss of generality that $[F_r(\mathbf x)]_q \le [F_s(\mathbf y)]_q$.
Then because production functions are $L$-Lipschitz, the gradient is bounded by $L$ and so we have that,
\[
    \mu\bigl([F_s(\mathbf y)]_q-[F_r(\mathbf x)]_q\bigr)
    \le m_{r,q}(\mathbf x)^2 \leq L^2.
\]
Here, the latter term comes from the production functions being $L$-Lipschitz. 
Rearranging gives that 
\[
    \bigl|[F_r(\mathbf x)]_q-[F_s(\mathbf y)]_q\bigr|
    \le \frac{L^2}{\mu}.
\]
Therefore, $|\theta_{i,q}-\theta_{r,q}| \leq L^2/\mu$, and so 
\[
    p_{i,q}
    :=
    \frac{w_q}{N-1}
    \sum_{r\neq i}
    \sigma'\!\left(\theta_{i,q}-\theta_{r,q}\right)
    \ge
    w_{\min}
    \min_{|z|\le L^2/\mu}\sigma'(z)
    =
    \underline p.
\]

Next, let $s_i:=\|\mathbf x_i\|_1$. 
We consider two cases depending on whether or not the trailing developer $i$ surpasses developer $j$ at timestep $t+1$. 
In the first case, suppose $\theta_{i,k}^{t+1}\ge \theta_{j,k}^{t+1}$, then 
\[
    \Delta\theta_{i,k}^t
    =
    \theta_{i,k}^{t+1}-\theta_{i,k}^t
    \ge
    \theta_{j,k}^{t+1}-\theta_{i,k}^t \geq \theta_{j,k}^{t}-\theta_{i,k}^t
    =
    g_{i,k}^t.
\]
Here, the last inequality comes from the fact that investments are non-negative and production functions are non-decreasing. 
As a result, we have that $ \Delta\theta_{i,k}^t\ge \beta g_{i,k}^t$ for any $\beta \leq 1$. 

For the second case, suppose that $\theta_{i,k}^{t+1}<\theta_{j,k}^{t+1}$.
Then we can bound this difference by noting that $\mu (\theta_{j,k}^{t+1}-\theta_{i,k}^{t+1}) \leq m_{i,k}(\mathbf{x})^2$. 
We bound $m_{i,k}(\mathbf{x})$ by noting that at equilibrium, we have that $c'(s_i)=\sum_{q=1}^K p_{i,q}\frac{\partial [F_i(\mathbf X_i^{t+1})]_q} {\partial x_\ell}$, and so 
\[
    c'(s_i)
    \ge
    p_{i,k}
    \frac{\partial [F_i(\mathbf X_i^{t+1})]_k}
         {\partial x_\ell}
    \ge
    \underline p\,
    m_{i,k}(\mathbf X_i^{t+1}).
\]
We note that $\theta_{j,k}^{t+1}-\theta_{i,k}^{t+1} > 0$ and so $ m_{i,k}(\mathbf X_i^{t+1})^2 \geq \mu (\theta^{t+1}_{j,k}-\theta^{t+1}_{i,k}) > 0$, and so zero investment cannot satisfy the first-order optimality condition; hence $s_i>0$ and such an
$\ell$ exists.
We next bound $c'(s_i)$ as 
\[
    c'(s_i)
    =
    \int_0^{s_i}c''(u)\,du
    \le Ms_i.
\]
Therefore
\[
Ms_{i} \geq c'(s_{i}) \geq \underline{p} m_{i,k}(\mathbf{X}_{i,}^{t+1}).
\]
Rearranging gives 
\[
    s_i
    \ge
    \frac{\underline p}{M}
    m_{i,k}(\mathbf X_i^{t+1}).
\]
We finally substitute this back in and simplify with $\Delta \theta^{t}_{i,k}$ using the concavity of $F_{i}$ to get 
\[
\begin{aligned}
    \Delta\theta_{i,k}^t
    &=
    [F_i(\mathbf X_i^{t+1})]_k
    -
    [F_i(\mathbf X_i^t)]_k\\
    &\ge
    \nabla [F_i(\mathbf X_i^{t+1})]_k^\top
    \mathbf{x}_i \\
    &\ge
    m_{i,k}(\mathbf X_i^{t+1})\,s_i \\ 
    &\ge 
    \frac{\underline p}{M}
    m_{i,k}(\mathbf X_i^{t+1})^2 \\ 
    &\ge 
    \frac{\mu \underline p}{M} (\theta_{j,k}^{t+1}-\theta_{i,k}^{t+1}) \\ 
    &\ge 
    \frac{\mu \underline p}{M} (g_{i,k}^t-\Delta \theta_{i,k}^t).
\end{aligned}
\]
Here, the last inequality comes from $\theta_{j,k}^{t+1}-\theta_{i,k}^{t+1} = 
    g_{i,k}^t
      +\Delta\theta_{j,k}^t
      -\Delta\theta_{i,k}^t \ge
    g_{i,k}^t-\Delta\theta_{i,k}^t$.
Rearranging gives 
\[
    \Delta\theta_{i,k}^t
    \ge
    \frac{\underline p\mu}
         {M+\underline p\mu}
    g_{i,k}^t
    =
    \beta g_{i,k}^t.
\]
\end{proof}